\documentclass{article}
\usepackage[a4paper, total={6in, 10in}]{geometry}

\usepackage{cite}

\usepackage{graphicx,comment} % Required for inserting images
\usepackage{appendix}
\usepackage{booktabs}
\usepackage{blindtext}
\usepackage{graphicx}
\usepackage{multirow}
\usepackage{tabularx}
\usepackage{subfig}
\usepackage{caption}
\usepackage{appendix}
\usepackage{hyperref}% add hypertext 
\hypersetup{colorlinks=true, citecolor=blue, urlcolor=blue, linkcolor=blue}
\usepackage{color}
\usepackage{authblk}

\usepackage{subfig}
\usepackage{verbatim}
\usepackage{diagbox}
\usepackage{amsmath} 
\usepackage{siunitx}
\usepackage{threeparttable}
\usepackage{adjustbox}
\usepackage[figuresleft]{rotating}
\usepackage{enumitem}

\title{Beyond a binary theorizing of prosociality}

\author[a]{Chen Shen}
\author[b,e]{Zhixue He}
\author[c]{Hao Guo}
\author[d]{Shuyue Hu}
\author[a,e]{Jun Tanimoto}
\author[b,1]{Lei Shi\thanks{shi\_lei65@hotmail.com}}
\author[f,g]{Petter Holme\thanks{pttrhlm@gmail.com}}

\affil[a]{Faculty of Engineering Sciences, Kyushu University, Fukuoka 816-8580, Japan}
\affil[b]{School of Statistics and Mathematics, Yunnan University of Finance and Economics, Kunming 650221, China}
\affil[c]{Department of Computer Science and Technology, Tsinghua University, Beijing 100084, China}
\affil[d]{Shanghai Artificial Intelligence Laboratory, Shanghai, China}
\affil[e]{Interdisciplinary Graduate School of Engineering Sciences, Kyushu University, Fukuoka 816-8580, Japan}
\affil[f]{Department of Computer Science, Aalto University, Espoo 02150, Finland}
\affil[g]{Center for Computational Social Science, Kobe University, Kobe 650-0017, Japan}

\date{\today}

\begin{document}

\maketitle

\begin{abstract}
A stylized experiment, the public goods game, has taught us the peculiar reproducible fact that humans tend to contribute more to shared resources than expected from economically rational assumptions. There have been two competing explanations for this phenomenon: either contributing to the public good is an innate human trait (the prosocial preference hypothesis) or a transitory effect while learning the game (the confused learner hypothesis). We use large-scale experimental data from a novel experimental design to distinguish between these two hypotheses. By monitoring the effects of zealots (persistently cooperating bots) and varying the participants' awareness of them, we find a considerably more complex scenario than previously reported. People indeed have a prosocial bias, but not to the degree that they always forego taking action to increase their profit. While our findings end the simplistic theorizing of prosociality in the public goods game, an observed positive, cooperative response to zealots has actionable policy implications.
\end{abstract}

\clearpage
\section{Introduction}

Public goods, such as environmental conservation and public health initiatives, rely on collective contributions for their establishment and maintenance, presenting a challenge due to the potential for free-riding~\cite{kaul2003providing,ledyard1994public}. The public goods game serves as a key tool in exploring the behavioral aspects of free-riding and how to mitigate it. In this game, participants decide whether to contribute part of their endowment to a common pool, which is then distributed equally among all participants. Despite the theoretical incentive for non-contribution~\cite{archetti2012game,anderson1998theoretical}, empirical studies consistently reveal that a majority of individuals do contribute initially, with contributions stabilizing at lower levels over time~\cite{chaudhuri2011sustaining,fehr2018normative}. These observations have given rise to two primary theories: the `prosocial preferences' hypothesis~\cite{camerer2006does,fehr2018normative,norenzayan2008origin,fehr2003nature} and the `confused learners' hypothesis~\cite{burton2013prosocial,burton2016conditional,andreoni1995cooperation,houser2002revisiting,nielsen2022sharing}. These hypotheses offer differing perspectives on the motivations that drive human cooperation, shaping the ongoing debate~\cite{burton2013prosocial,burton2016conditional,burton2022restart,burton2021payoff,fischbacher2010social,camerer2013experimental,wang2024confusion} between altruism and self-interest in the context of public goods games.

The prosocial preferences hypothesis suggests that individuals vary in their valuation of others' welfare, with a majority being altruistically motivated, keen to contribute, and averse to unfair outcomes, while a minority are selfish free riders who contribute nothing. Initially, prosocial individuals often have optimistic expectations about others' cooperation. However, as the game proceeds, they perceive or believe that their contributions are being exploited by free riders, and thus reduce their contributions to prevent inequity and promote fairness, leading to a decline in overall group contributions~\cite{fehr2003nature,fischbacher2010social}.

The confused learners hypothesis posits that individuals inherently prioritize self-interest, and it is an incomplete understanding of how to maximize profit that leads to cooperative behavior. This incompleteness may stem from uncertainties or misconceptions about the game’s rules, including the costs of contributing. As individuals gain experience with the game, they gradually recognize that free-riding is a more advantageous strategy, resulting in decreased contributions and a consequent decline in overall group contributions~\cite{andreoni1995cooperation,houser2002revisiting,burton2013prosocial,burton2016conditional}.

Previous research into these mechanisms has varied the amount of information available about other participants and their performance. 
Research promoting the confused learners hypothesis has relied on mixing human and computer players, where cooperative play against computers has been interpreted as not representing prosociality~\cite{houser2002revisiting,burton2016conditional}. In a related experimental design, the \textit{black box method}, participants were not even informed that they played the game with other people~\cite{burton2013prosocial,burton2022black}. Since there were only small differences between experiments where the concern for others had been eliminated and not, the authors concluded that something other than prosociality, i.e., the confused learners hypothesis, explains the high cooperation. However, others have criticized the conclusions from these experiments. Ref.~\cite{camerer2013experimental} suggests that a perceived reciprocity of the black box explains the above small differences. Ref.~\cite{wang2024confusion} give more information and time to the players before the experiment but don't observe the consistency in cooperation that the confused learner hypothesis predicts in that situation.

The key ingredient in our experimental design is the presence of zealots---players who unconditionally contribute to the social good. The prosocial preferences hypothesis predicts a greater cooperation in the presence of zealots. 
% HERE
This anticipation stems from the assumption that individuals, when interacting with zealots, will have optimistic expectations about others' cooperation due to the certainty of their contributions being matched. In contrast, the confused learners hypothesis anticipates consistent cooperation levels across both conditions. According to this hypothesis, interactions with zealots would not affect individuals' understanding of how to maximize their personal gains, leading to similar levels of cooperation regardless of zealot presence.

Employing the human-computer interaction method allows us to examine whether a reduction in concern for others' interests affects behavior. This method involves informing participants that the zealots are actually bots, incapable of deriving any material gain from human players. If the level of cooperation remains unchanged when participants are informed of this fact, compared to when they are not, it would lend support to the confused learners hypothesis, suggesting that motivations related to concern for others have minimal impact on behavior. Conversely, if informing participants that zealots are bots leads to a decrease in cooperation levels, it suggests that prosocial preferences are at play, as participants adjust their behavior based on their reduced concern for the altruistic motives of others.

Opting for the prisoner's dilemma game, a two-player variant of the public goods game~\cite{hauert2003prisoner}, streamlines our experiment by focusing on binary decisions of cooperation or defection while bypassing complexities associated with quantifying zealot contributions in general public goods games. Anticipating a decrease in average group contributions based on existing literature~\cite{chaudhuri2011sustaining,li2018punishment}, we incorporate three treatments: a control scenario without zealots, where players interact solely with other human players; an unaware-of-zealots scenario, where players interact with zealots without being informed of their opponents' zealotry; and an informed-of-zealots scenario, where players are explicitly informed of their opponents' zealotry. Participants experience each treatment twice, in accordance with the Latin square design that counters order effects~\cite{kirk2009experimental}. We then supplement the study results using a human-computer interaction method, including a between-subjects treatment with zealot bots in which participants are informed about facing bots who are committed to cooperation and who receive no material gain from human players.

\section{Results}

\begin{figure}[!t]
    \centering
{\includegraphics[width=0.96\linewidth]{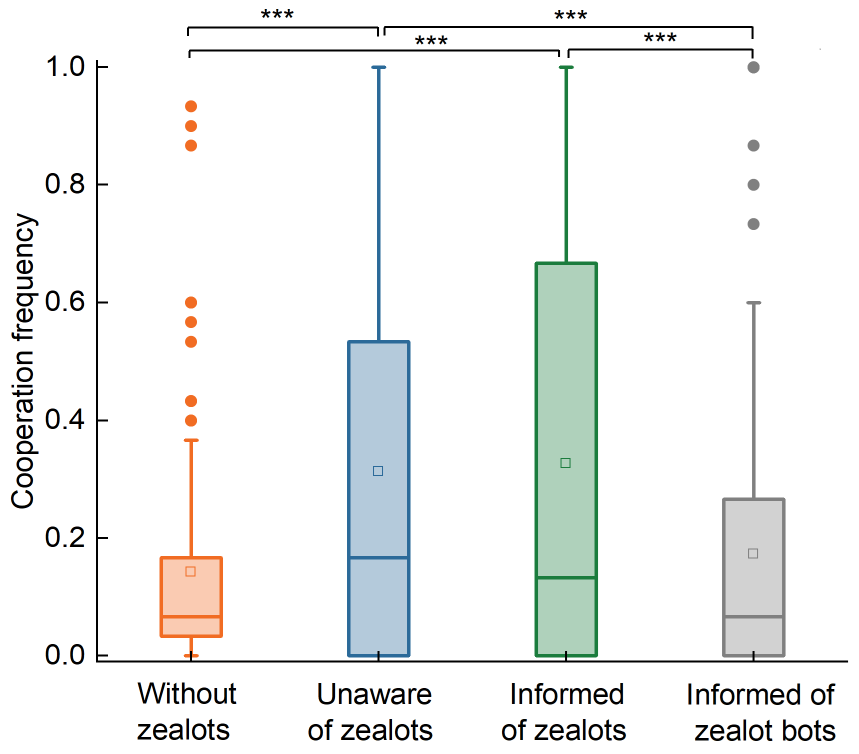}}
    \caption{\textbf{The presence of zealots stimulates cooperativeness.} The average cooperation frequency of 14.3\% in the treatment without zealots is significantly lower (pairwise $t$ test; $t=-6.906$, $p$-value < $10^{-10}$) than the 31.3\% in the unaware-of-zealots treatment, and also significantly lower (pairwise $t$ test; $t=-6.638$, $p$-value < $10^{-10}$) than the 32.8\% in the informed-of-zealots treatment. Moreover, no significant difference in average cooperation frequency was found between the unaware and informed-of-zealots treatments (pairwise $t$ test; $t=-0.969$, $p$-value = $0.334$). However, the average cooperation frequency of 17.3\% in the informed-of-zealot-bots treatment was significantly lower than in both the unaware and informed-of-zealots treatments (two sample $t$-test; $t=-3.67$, $p$-value = $0.0003$ and $t=-3.96$, $p$-value = $0.0002$, respectively), and no significant difference was observed compared to the treatment without zealots (two sample $t$-test; $t=1.12$, $p$-value = $0.26$). All these results are further examined using ANOVA (Analysis of Variance, see Tables S6 and S7). Boxplots illustrate the empirical distribution of cooperation frequencies, calculated by dividing the number of cooperative actions by the total number of rounds for each participant. The height of the box represents the interquartile range, the horizontal line indicates the median, and the square within the box marks the mean. Points outside this range are identified as outliers.}
        \label{fig1}
\end{figure}

Briefly, from a methodological standpoint, we ensured that the observed effects were solely due to the presence of zealots by using the one-shot and anonymous prisoner's dilemma game. Each treatment comprised 15 rounds, maintaining strict subject anonymity throughout, and no subject was ever paired with the same subject more than once. This setup prevented potential influences of reciprocity~\cite{trivers1971evolution,nowak1998dynamics}, reputation~\cite{nowak2005evolution}, or strategic cooperation~\cite{kreps1982rational}. The game was based on a payoff matrix such that action 1 ($C$) entailed giving up one unit for the opponent to receive three units, while action 2 ($D$) meant earning one unit at the opponent's expense of one unit. For robustness, we repeated the experiment with another payoff matrix, and with several player-to-zealot ratios. For this, we recruited a total of 964 participants, with 122 engaging in each within-subjects treatment and 120 in each between-subjects treatment involving zealot bots.  The study received approval from the Yunnan University of Finance and Economics Ethics Committee, and all participants provided informed consent. Methodological details are provided in Materials and Methods and Supporting Information (SI) Supporting Methods.

\subsection*{\bf{Zealots stimulate cooperativeness}}

In the one-shot and anonymous prisoner's dilemma game, the presence of zealots promotes cooperation (Figure~\ref{fig1}). In comparison to the control treatment, cooperation frequencies are significantly higher in treatments involving zealots, regardless of whether participants are informed about their opponents' zealotry. Interestingly, we found no significant difference in cooperation frequencies between participants who were informed about zealots and those who were not. However, when participants were explicitly informed that their opponents were zealot bots, a significant decrease in cooperation frequency was observed. This decrease aligns closely with the control treatment.

\begin{figure}[!t]
    \centering
{\includegraphics[width=0.98\linewidth]{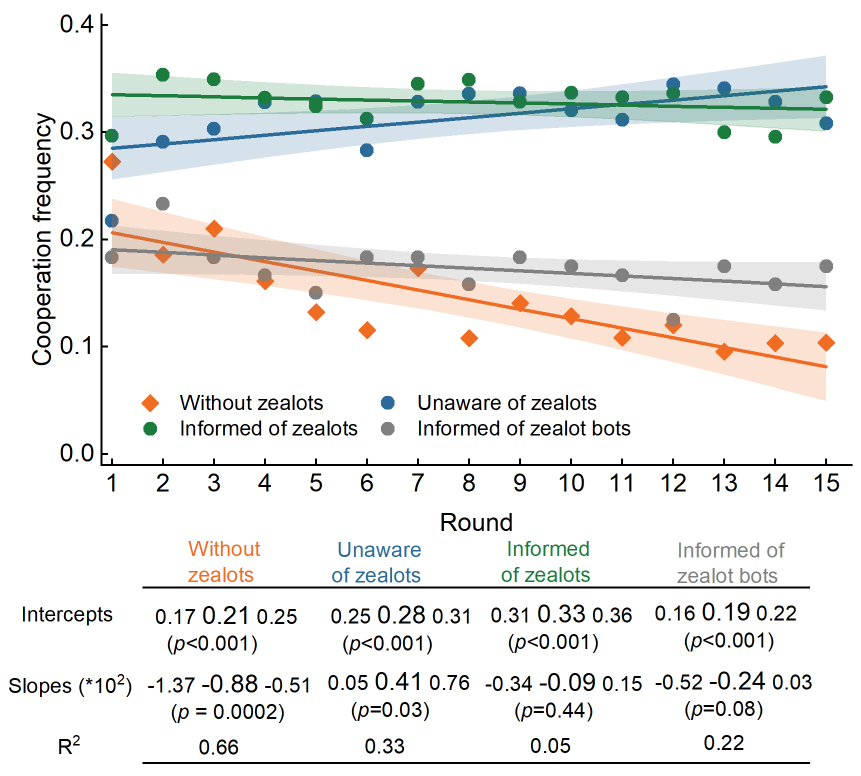}}
    \caption{The presence of zealots stabilizes cooperation. Shown are the frequency of cooperation over time, illustrated by data points and regression lines for four distinct treatments: without zealots, unaware of zealots, informed of zealots, and informed of zealot bots. In the treatment without zealots, the frequency of cooperation exhibits a slowly decreasing trend, with a slope significantly different from zero. In contrast, treatments involving zealots---regardless of whether participants were aware of their opponents' zealotry or in the informed-of-zealot-bots scenario---demonstrate either a slight increase or stability in cooperation levels. This suggests that the presence of zealots contributes to stabilizing cooperation.  The frequencies were calculated by dividing the number of participants who chose to cooperate by the total number of participants in each round. Smaller font sizes and shaded areas represent 95\% confidence intervals.}
        \label{fig2}
\end{figure}

Further analysis was conducted to validate these findings and explore cooperation trends over time (Figure ~\ref{fig2}). The frequency of cooperation exhibits a slowly decreasing trend in the treatment without zealots, with a slope significantly different from zero. In contrast, introducing zealots significantly alters the trend such that the frequency of cooperation now shows a slowly increasing trend in the treatment where participants are unaware of zealots or remains stable in the treatments where participants are informed of zealots or informed of zealot bots. All this indicates that the presence of zealots has a stabilizing impact on cooperation.

Initial and final cooperation levels were examined for each treatment (SI, Table S3). There are no significant differences in initial cooperation levels between treatments with and without zealots, regardless of whether participants are aware of zealots' presence. However, significant differences are observed in the final-round cooperation levels among these treatments. In treatments where participants are unaware or informed of zealots, the final cooperation levels are significantly higher than in the treatment without zealots. Conversely, in the treatment with participants informed about zealot bots, the final cooperation level shows no significant difference compared to the treatment without zealots.

The described results pertain to the data pooled across three sessions. However, the order in which treatments were presented to participants differed between sessions (SI, Figure S5), offering additional insights into participant behaviors. Notably, the presence of zealots can restart and maintain higher levels of cooperation, even after experiencing low cooperation levels in the treatment without zealots. This immediate effect occurs regardless of participants' awareness of their opponents' zealotry. Interestingly, when participants engaged in the treatment without zealots after experiencing treatments involving zealots, the average frequency of cooperation generally decreased over time, although there was one exception where an elevated level of cooperation was maintained.

Overall, the results suggest that the presence of zealots in prisoner's dilemma games significantly increases and maintains cooperation levels, contradicting the confused learner hypothesis. While prosocial preferences play a role, participants do not match the cooperativeness of their zealot opponents nearly perfectly, indicating that individuals are not solely motivated by concerns for others.

\subsection*{\bf{The role of social value orientations}}

\begin{figure*}[!t]
    \centering
{\includegraphics[width=0.71\linewidth]{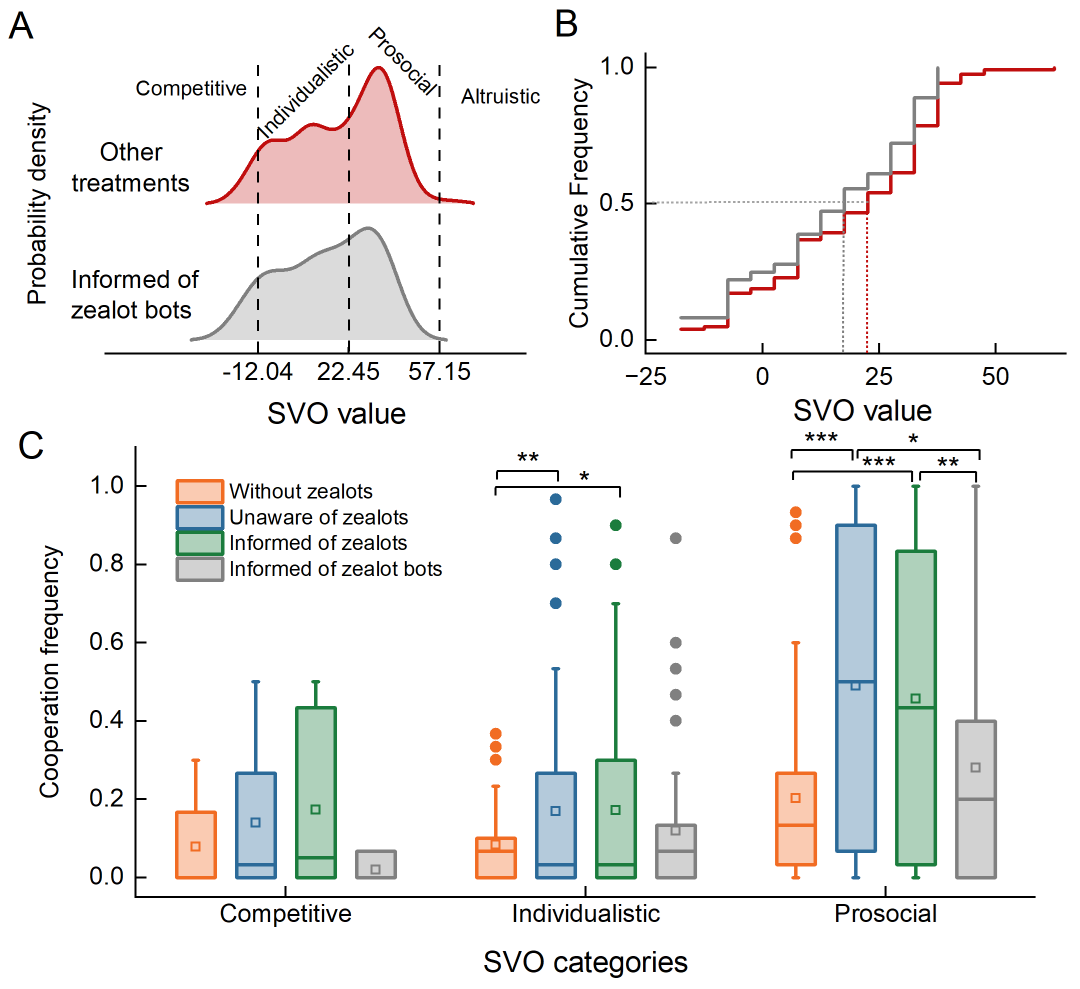}}
    \caption{\textbf{The impact of social value orientation on cooperation across treatments.} A. Probability density plots illustrate the distribution of participants' social value orientations (SVO)—classified as competitive, individualistic, prosocial, and altruistic based on the scheme by ref.~\cite{murphy2011measuring}. These plots show the SVO distributions for two groups. The first group experienced a within-subjects design with three treatments: without zealots, unaware-of-zealots, and informed-of-zealots. The second group participated only in the informed-of-zealot-bots treatment. These distributions are visualized through smoothed kernel density estimation. B. Cumulative distribution functions of SVO scores reveal a significantly lower median value in the informed-of-zealot-bots treatment compared to the other treatments (median values 15.2 vs.\ 22.5; two-tailed Mann–Whitney $U$ test; $W=6027.5$, $p=0.02$). C. Boxplots show cooperation frequencies across SVO categories in four treatments. Competitive categories exhibit consistent cooperation frequencies across treatments. Individualistic participants increased cooperation in treatments involving unaware (17.1\%) and informed of zealots (17\%) compared to without zealots (8.4\%; pair-wise $t$ tests: $t=-3.04$, $p=0.004$; $t=-2.59$, $p=0.01$), with no significant difference compared to informed of zealot bots (11.9\%; two-sample $t$-test: $t=-1.41$, $p=0.16$; $t=-1.24$, $p=0.22$). No significant difference was found between treatments without zealots and informed of zealot bots (two-sample $t$-test; $t=1.43$, $p$-value = 0.16). Prosocial participants significantly increased cooperation in the treatments involving unaware and informed of zealots (44.7\%, 48.1\%) versus without zealots (19.1\%; pair-wise $t$ tests: $t=6.5$, $p<10^{-8}$; $t=6.6$, $p<10^{-8}$) and informed-of-zealot-bots treatment (28\%; two-sample $t$-test: $t=-2.51$, $p=0.01$; $t=-2.961$, $p=0.004$), with no significant difference between treatments without zealots and informed of zealot bots (two-sample $t$-test: $t=1.71$, $p=0.09$).}
        \label{fig3}
\end{figure*}

To further understand the impact of zealots on cooperation, we assessed participants' social value orientations (SVO) post-experimentally and analyzed their payoff distributions. This helps us identify which orientations respond more to increased cooperation in the presence of zealots and whether this cooperation is driven by fairness, equality, or strategic self-interest.

Participants were categorized into four groups---competitive, individualistic, prosocial, and altruistic---following established classification criteria~\cite{murphy2011measuring} (see Table S5). Due to the limited number of participants classified as altruistic, they were combined with those identified as prosocial for analysis purposes. 

\begin{figure*}[!t]
    \centering
{\includegraphics[width=0.96\linewidth]{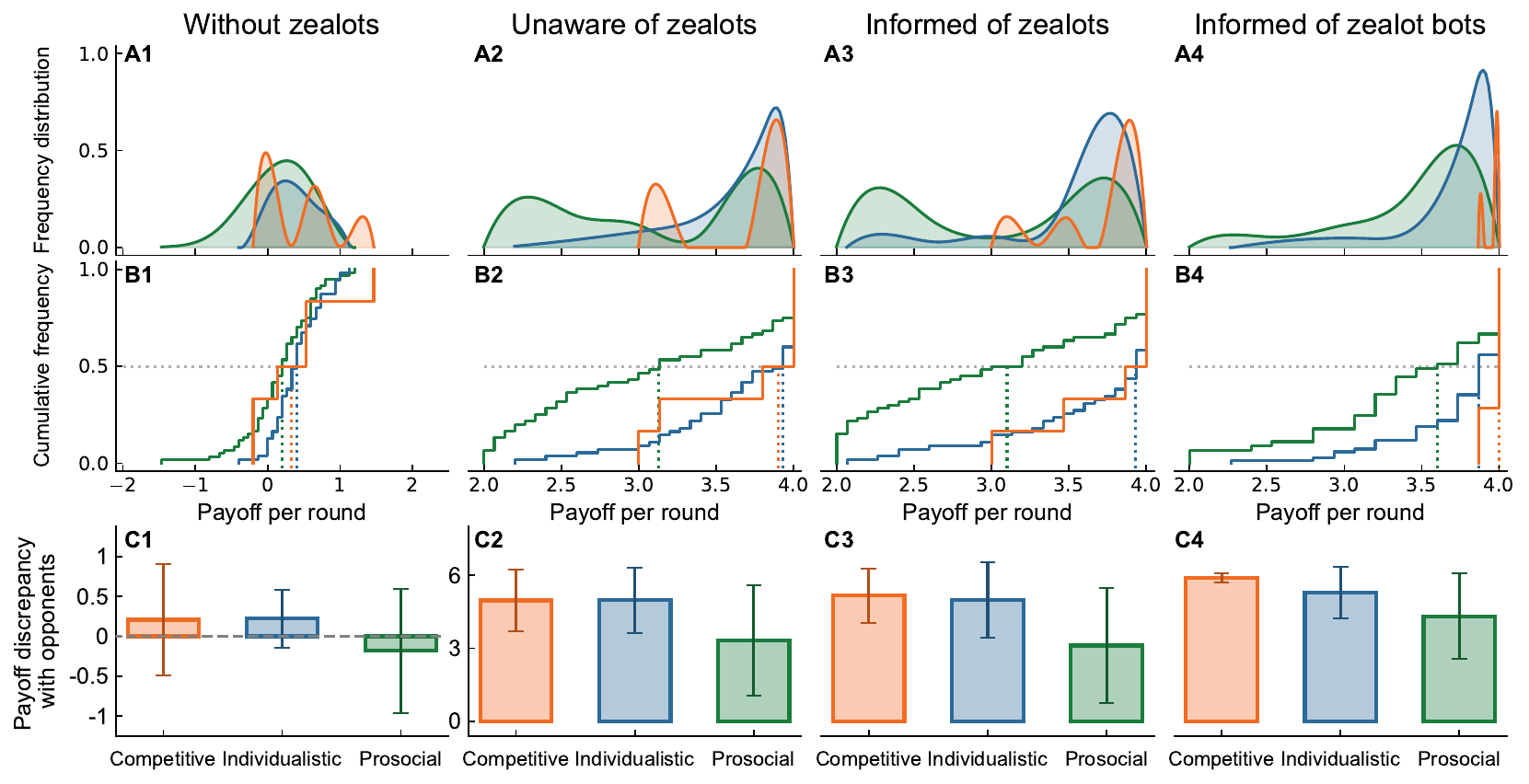}}
    \caption{\textbf{Comparative analysis of payoffs reveals prosocial preference with self-interest tendencies.} A. Probability density plots distinguish the distributions of payoff per round among participants identified as competitive, individualistic, and prosocial across four treatments: without zealots, unaware of zealots, informed of zealots, and with informed zealot bots, from left to right, respectively. B. Cumulative density comparisons highlight differences in these distributions across treatments. In the without-zealots treatment, payoffs of individuals with a competitive orientation (median value 0.33) show no significant difference compared to those with an individualistic (median values 0.4, Mann–Whitney U test: 144, $p=0.618$) or prosocial orientation (median values 0.2, Mann–Whitney U test: 194, $p=0.763$). Payoffs between individuals with individualistic and prosocial orientations show significant differences (Mann–Whitney U test: 2106, $p=0.011$). Introducing zealots or zealot bots, no notable differences in payoffs between competitive and individualistic categories are observed under each treatment (Median values 3.9, and 3.93, Mann–Whitney U: 169, $p=0.930$ in the unaware-of-zealots treatment; Median values 3.93 and 3.93,  Mann–Whitney U test: 173.5, $p=0.84$ in the informed-of-zealots treatment; Median values 4.0 and 3.867, Mann–Whitney U: 327, $p=0.087$ in the informed-of-zealot-bots treatment). However, prosocial category's median payoff values in treatments of unaware, informed of zealots, and informed zealot bots (median values 3.13, 3.1, 3.6) are significantly lower compared to those of competitive (Mann–Whitney $U$ test: 140.5, $p=0.041$; 90.5, $p=0.044$; 69.5, $p=0.016$) and individualistic categories (Mann–Whitney U test: 1002.5, $p<0.001$; 925, $p<0.001$; 1074, $p=0.006$). C. Bar charts reveal that participants generally earn more than their opponents across all treatments, except in the without-zealots treatment where competitive participants' payoffs (one sample $t$-test, mean value 0.333, $t=0.663$, $p=0.536$) and prosocial participants' payoffs (one sample $t$-test, mean value $-0.293$, $t=-1.814$, $p=0.075$) do not significantly differ from their opponents, individualistic participants' payoffs are significantly higher (one sample $t$-test, mean value 0.352, $t=4.472$, $p<0.001$). In the treatments of unaware of zealots, informed of zealots, and informed zealot bots, the differences in payoffs between participants and their opponents do not significantly alter for individualistic (median values 5.7, 5.8 and 6.0, Kruskal-Wallis test: $H=1.774$, $p=0.412$ ) and competitive players (median values 5.8, 5.8 and 5.6, $H= 0.877$, $p=0.645$). However, for prosocial players, the difference in payoffs increases significantly in the zealot bots scenario (median value 4.8) compared to the other two (median values 3.4, 3.3, Kruskal-Wallis test: $H=6.681$, $p=0.035$).
 }
        \label{fig4}
\end{figure*}

Analysis of SVO scores (Figure~\ref{fig3}B) shows a decline in median scores in the informed-of-zealot-bots treatment compared to treatments with human counterparts or undisclosed zealot bots.  This suggests that knowing the zealots are non-materialistic bots dampens prosocial tendencies. While cooperation frequencies among competitive and individualistic participants generally do not significantly differ across treatments (see SI, Table S8), individualistic participants exhibited an increase in their cooperation levels in treatments involving zealots compared to those without zealots (Figure~\ref{fig3}C). Prosocial participants increased cooperation in treatments with zealots. However, when informed their opponents were zealot bots, prosocial participants decreased their cooperation to levels observed in treatments without zealots (see also Table S9).

In the absence of zealots, competitive, individualistic, and prosocial participants  exhibit similar payoff distributions and median values (Figure \ref{fig4}). However, with zealots introduced, prosocial participants experience significantly lower payoffs compared to competitive and individualistic participants, regardless of whether zealot status is disclosed. Interestingly, when participants are informed of zealot bots' non-materialistic nature, prosocial participants' median payoffs improve but remain lower than those of competitive and individualistic participants. Moreover, prosocial participants consistently achieve higher payoffs than their zealot opponents across treatments, indicating both altruistic tendencies and some self-interest motivations.

Overall, these results suggest that competitive and individualistic participants remain self-interested, as they can obtain near-maximum payoffs when playing with zealots. Prosocial participants, on the other hand, reflect a preference for reciprocity but also show signs of nuanced strategic concerns when the benefits of prosociality are diminished.

\subsection*{\bf{Robustness}}
To validate the robustness of our findings, we conducted additional experiments using a different payoff matrix. In this matrix, cooperation incurred a cost of one unit to confer a benefit of two units to the partner, compared to the 3-unit benefit originally. Despite this variation, defection remained the strict Nash equilibrium in both cases, ensuring the preservation of the essential social dilemma of the game.

A comparative analysis shows that the results using the modified payoff matrix closely mirror the original results (Figures S6-S9). This reaffirms our findings that individuals exhibit other-regarding preferences but lean toward self-interest under certain circumstances.

\begin{figure}[!t]
    \centering
{\includegraphics[width=0.81\linewidth]{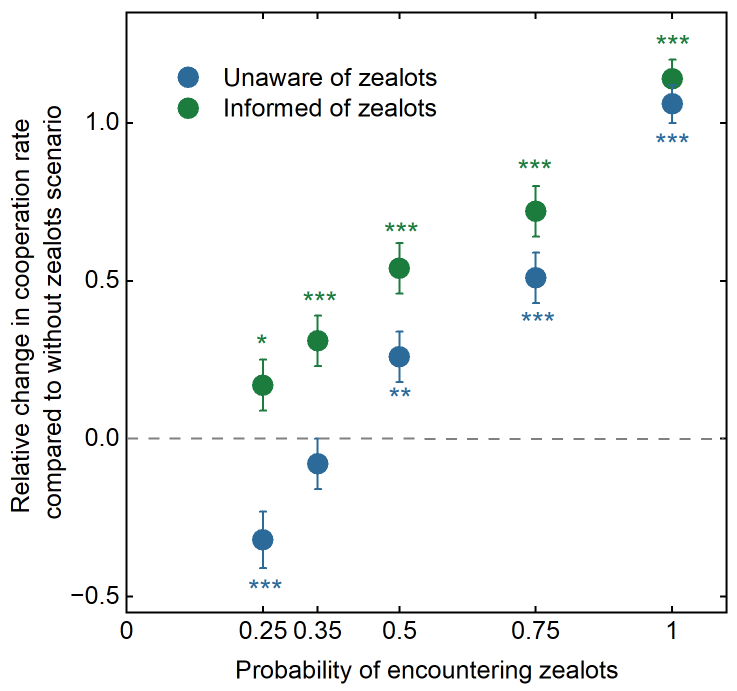}}
    \caption{Minority of zealots enhances cooperation, especially when participants are informed about their opponents' zealotry. Relative to the baseline condition without zealots (encounter probability $q= 0$), the influence of zealots on cooperation varies significantly based on participants' awareness.  In the informed-zealots treatments, encountering zealots at a low probability of 0.25 significantly promotes cooperation (coefficients of 0.17, $p$-value $= 0.04$), with increasing effects at probabilities of 0.35, 0.5, 0.75, and 1 (coefficients of 0.31, 0.54, 0.72, and 1.14, respectively, and $p$-value $< 10^{-5}$ for these scenarios). Conversely, in the unaware-of-zealots treatments, the probability of 0.35 shows no significant effect on cooperation  (coefficients of $-0.08$ $p$-value $= 0.34$). However, at higher probabilities of 0.5, 0.75, and 1, cooperation is significantly enhanced (coefficients of 0.26, 0.4, and 1.08 for probabilities of 0.5, 0.75, and 1, respectively; $p$-value $=0.001$ at probability of 0.5 and  $p$-value  $< 10^{-7}$ for others). Notably, a potential cooperation-diminishing effect is observed at a probability of 0.25 when participants are unaware of zealots (coefficient of $-0.32$, $p$-value = $0.0004$). These estimates are derived from generalized linear regression models (refer to SI Table S14). Vertical bars represent standard errors. Significance: $^\ast$ $p<0.05$ and $^{\ast \ast \ast}$ $p<0.001$. }
        \label{fig5}
\end{figure}

\subsection*{\bf{Leveraging prosocial tendencies}}

The results robustly demonstrate that people inherently possess prosocial preferences, though the valuation of others' welfare varies. This finding raises the question of how to effectively leverage this trait to enhance cooperation. So far, our analysis involved an equal number of participants and zealots, leaving it undetermined whether a minority of zealots could stimulate cooperation. To address this gap, we conducted additional experiments varying the probability $q$ that participants would encounter zealots. These experiments were conducted under conditions where participants were unaware and informed about their opponents' zealotry. Details of the experimental setup can be found in the Materials and Methods section and SI Supporting Methods. We chose encounter probabilities of 0.25, 0.35, 0.5, and 0.75, determined via the bisection method. Using generalized linear regression models, we assessed the influence of zealots on cooperation compared to the treatment without zealots, which served as a baseline reference in our models (see SI Table S14).

The findings, illustrated in Figure~\ref{fig5}, reveal that a minority of zealots can promote cooperation. This effect was more pronounced when participants were aware of their opponents' zealotry than when this information was not disclosed. In the informed-of-zealots treatment, the threshold of the cooperation-enhancing effect of zealots lies within an encounter probability range of [0, 0.25]. Conversely, in the treatment where participants were unaware of zealots, the threshold for a cooperation-promoting effect falls within [0.35, 0.5]. Notably, a negative effect on cooperation was detected at a lower encounter probability in the unaware-of-zealots treatment, suggesting that the impact of zealots on cooperation might be dual.

\section{Discussion}
Our study aimed to explore the impact of introducing zealots—players who unconditionally cooperate—on cooperation levels in strict one-shot anonymous prisoner's dilemma games. We varied the information given to participants about their zealot opponents to investigate two main hypotheses: the prosocial preferences hypothesis, which posits that individuals are motivated by fairness and altruism, and the confused learners hypothesis, which suggests that higher-than-expected cooperation levels stem from misunderstanding the game's rules. Our findings challenge existing theories by revealing that neither fully explains the observed cooperation and suggesting that human cooperation is far more complicated than previously understood, shaped by mixed motives of altruism and selfishness.

\paragraph{\bf{Prosocial preferences and confusion.}}

Our findings cannot be fully explained by either hypothesis. Specifically, our results challenge the prosocial preferences hypothesis in several ways. First, cooperation levels with zealot opponents were significantly lower than 100\%, even when prosocial participants were informed about the zealots' guaranteed cooperation. This is inconsistent with predictions from theories of prosocial behavior, which suggest that prosocial preferences could transform a prisoner's dilemma into an assurance game, where mutual cooperation would be a stable equilibrium if individuals believe that their opponents will also cooperate ~\cite{fehr2003nature,fehr1999theory,bolton2000erc,falk2001distributional}. According to these theories, individuals motivated by fairness and altruism should reciprocate the guaranteed cooperation of zealots, leading to near-perfect cooperation levels. The lower cooperation levels observed in our study suggest that other factors are influencing participants' decisions. 

Second, participants gained significantly higher payoffs than their zealot opponents. If inequity aversion was the primary motivator, participants would be uncomfortable with gaining significantly more than their opponents. This discrepancy suggests that participants are not only driven by concerns for fairness or equality, which are central to prosocial preferences.

%Third, no significant difference in cooperation levels was found between informed and uninformed treatments regarding opponents' zealotry in scenarios where participants were certain to play with zealots. 

Third, no significant difference in cooperation levels was found between the unaware-of-zealots treatment and the informed-of-zealots treatment. If prosocial preferences were primarily belief-driven, informing participants about their opponents' cooperative nature should have resulted in higher cooperation levels. The lack of significant difference indicates that participants' cooperation decisions are not strongly influenced by their knowledge of opponents' guaranteed cooperation. Although we observed a significant difference when participants were not certain to be matched with zealot opponents (refer to SI, Figure S10), this indicates that prosocial preferences are indeed belief-driven under certain conditions. However, the overall explanatory power of the prosocial preferences hypothesis remains limited when considering the full range of our findings.

%Third, no significant difference in cooperation levels was found between informed and uninformed treatments regarding opponents' zealotry. If prosocial preferences were primarily belief-driven, informing participants about their opponents' cooperative nature should have resulted in higher cooperation levels. The lack of significant difference indicates that participants' cooperation decisions are not strongly influenced by their knowledge of opponents' guaranteed cooperation. These findings suggest that the explanatory power of the prosocial preferences hypothesis is limited.

Fourth, the cooperation levels in the treatment without zealots and the treatment where participants were informed about the non-materialistic nature of their zealot bot opponents were consistent. If prosocial preferences were the primary motivator, informing participants that their opponents were bots should have led to a decrease in cooperation levels due to the reduced concern for the altruistic motives of others. The unchanged cooperation levels in these treatments suggest that prosocial preferences cannot fully explain this result, further challenging the prosocial preferences hypothesis.

Our results also challenge the confused learners hypothesis. First, the significant increase in cooperation levels when interacting with zealots compared to the control treatment indicates that participants can adjust their behavior based on the perceived cooperation of their opponents. This contradicts the notion that participants are simply confused about the game's payoff structure. Second, the significant difference in cooperation levels in treatments where participants were informed about the non-materialistic nature of their zealot bot opponents suggests that participants' decisions are not driven by confusion. According to the confused learners hypothesis, cooperation levels should be consistent across all treatments if participants were merely confused. The observed significant difference undermines this prediction. Third, participants' ability to gain significantly higher payoffs than their zealot opponents further undermines the confused learners hypothesis. This demonstrates their capacity to strategically optimize their outcomes rather than acting out of confusion. These findings indicate that participants' cooperation is informed by a nuanced understanding of the experimental setup, and not confused.

Our findings suggest that the reality of human cooperation is more complex than what is solely captured by the binary hypotheses of prosocial preferences and confused learners. While prosocial preferences argue that human cooperation is influenced by concerns for fairness and altruism, our results indicate that this is not the whole story. Our results reveal mixed motives of altruism and selfishness for the higher-than-expected cooperation. Although prosociality and selfishness are two important components at play, other factors such as strategic considerations, perceptions of authenticity, and social and psychological influences also play a role. For example, participants showed higher cooperation when the non-materialistic nature of zealot bots was undisclosed, indicating that perceptions of authenticity and genuine altruistic motives are crucial. Additionally, the ability of participants to gain higher payoffs than their zealot opponents suggests strategic considerations in optimizing outcomes. These factors highlight the need for a comprehensive model of human cooperation in social dilemma games. This model must account for the complexity and variability in human behavior, recognizing that sometimes behavior aligns with the confused learners hypothesis, sometimes with prosocial preferences, and sometimes with other motivations.

\paragraph{\bf{Implications}}

Although our results cannot be fully explained by prosocial preferences, the positive cooperative response to zealot opponents suggests important policy implications for enhancing cooperation in social dilemmas. Our study shows that a minority of zealots can significantly enhance cooperation, especially when their commitment is transparent to other players. This insight highlights the need to explore non-material incentives to foster human cooperation. Given humans' prosocial bias, government policies can emphasize the importance of altruism and ensure transparency of intentions through effective communication to nudge people toward cooperation. Additionally, the negative impact of low encounter probabilities of undisclosed zealots on cooperation suggests that carefully planning the introduction and visibility of zealots is essential and warrants further investigation.

While fostering optimistic beliefs about others' cooperation can enhance cooperation to some extent, this strategy alone is insufficient to fully encourage human cooperation. It is also crucial to provide material incentives that align cooperation with individuals' self-interest, given humans' mixed motives of altruism and selfishness. 

\paragraph{\bf{Conclusions}}
In conclusion, our study underscores the complexity of human cooperation, revealing that it is influenced by a blend of altruism and selfishness. Moving beyond binary hypotheses, future models of human cooperation must incorporate this complexity to predict and enhance cooperative behavior in diverse contexts effectively. By understanding and leveraging the motivational basis for cooperative behavior, we can design more effective interventions and policies that promote sustained cooperation in social and organizational settings.

The debate on these hypotheses has extended to scenarios involving altruistic punishment (punishing non-contributors at personal cost) and the restart effect of cooperation in repeated games (increased cooperation when players unexpectedly continue after the final round)~\cite{burton2022restart,burton2021decoupling,camerer2013experimental}. Repeated interactions introduce reciprocal altruism, where individuals sacrifice short-term benefits for long-term social gains, to explain prosociality. Altruistic punishment examines whether cooperation and punishment are linked altruistic traits or driven by self-interest. Our conclusions about cooperative behavior, drawn from one-shot, anonymous prisoner's dilemma games, may not address these complexities as they only consider confusion and prosocial preferences. Interacting with cooperating bots offers an alternative method to study the motivations behind prosociality in these complex scenarios. In the near future, studies should use this approach to explore repeated interactions, altruistic punishment, and varying anonymity levels, thus enhancing our understanding of human cooperation and guiding effective interventions. Further into the future we foresee research that closes the door our paper has opened---i.e., that models the human predisposition to cooperate beyond economic rationality, not just explains it in terms of simple causal relations.

%We conducted our experiments in a controlled laboratory setting using a student sample to ensure higher experimental control and better player understanding. The laboratory setting minimizes distractions and variability, providing consistency for all participants. While online experimental platforms such as Amazon Mechanical Turk offer a more diverse sample, they present challenges related to data quality and control. Using a student sample maintains high standards of quality and consistency, enhancing the reliability of our results. Future research should incorporate diverse samples to explore the generalizability of our findings in various contexts.

\section{Materials and Methods}
\subsection*{\bf{Experimental methods}}
We enrolled 964 undergraduate students from Yunnan University of Finance and Economics, Kunming, China, during October 2023 to March 2024. Participant demographics, including gender distribution and mean age, are detailed in the Supplementary Information (See also Table S2). All subjects took part in a one-shot prisoner's dilemma game held in the behavioral economics laboratory, School of Statistics and Mathematics. This facility houses around 100 computerized isolation cubicles running oTree software~\cite{chen2016otree}. To prevent interaction and ensure concentration, we maintained at least one empty cubicle between each occupied one. Experiment integrity was overseen by two supervisors who also managed technical issues.

The experimental design involved pooling players and randomly assigning them to one of twelve experimental treatments. The initial phase comprised six sessions split into two parts: a within-subjects design (122 volunteers) and an informed-of-zealot-bots treatment (120 volunteers). The within-subjects segment spanned three sessions, each containing three conditions—without zealots, unaware of zealots, and informed of zealots—designed to mitigate order effects via the Latin square method. Each condition was experienced twice per session. The informed-of-zealot-bots treatment also spanned three sessions, each with 40 participants. For robustness, this setup was repeated with an alternative payoff matrix and an additional 242 students, similarly distributed across six sessions. 

To assess the threshold of zealots necessary to promote cooperation, we used the bisection method, setting encounter probabilities at 0.25, 0.35, 0.5, and 0.75, thus creating four experimental conditions with a total of 480 volunteers. Each condition adopted a within-subjects design containing unaware of zealots and informed-of-zealots treatments, employed the Latin square method to counterbalance order effects, and included two sessions. All twelve experimental treatments were preregistered (\url{https://aspredicted.org/BSV_S1W}, and \url{https://aspredicted.org/F4P_JWX}).

Sessions began with the random allocation of volunteers to isolated computer cubicles where they read the on-screen instructions (see Supporting Methods in SI). This was followed by a pre-game test to check their understanding of the instructions (see Figure S1). Those who failed the test were asked to reread the instructions and retake the test. Afterward, participants engaged in random-order interactions to play the game (see SI, Figure S2 for the game interface). Depending on the experimental design, a session could contain multiple treatments (within-subjects design) or a single treatment (between-subjects design). The session concluded with a questionnaire assessing social value orientation (see SI, Figures S3 and S4). Each treatment consisted of 15 rounds, lasting approximately 30 minutes, with no participant attending more than one session. Volunteers started with an initial balance of 50 units, adjusted round-by-round based on decisions and payoff matrix rules. The final balance, if positive, was converted into a monetary payout at a rate of 0.5 Chinese Yuan (CNY) per unit. This was supplemented with a show-up fee of 15 CNY, resulting in an average payout of 82.09 CNY, ranging between 28 CNY and 180 CNY.

\paragraph{\bf{Statistical Analysis}}
We employed pairwise $t$-tests for within-subject treatment comparisons and independent sample $t$-tests for between-subject comparisons. Further, we utilized chi-square tests for contingency table analyses and two-way ANOVAs to examine the dependency of one continuous dependent variable on two categorical independent variables. All hypothesis tests were reported, including non-significant results. For trend identification over time, linear regression analyses were performed, and to determine the impact threshold of zealots on cooperation, generalized linear regression models were employed, with the without zealot treatment serving as the baseline.

\section*{Manuscript information}

\subsection*{\bf{Ethics statement}}
The experiment was approved by the Yunnan University of Finance and Economics Ethics Committee on the use of human participants in research, and carried out in accordance with all relevant guidelines. We obtained informed consent from all volunteers.

\subsection*{Acknowledgement}
We are grateful to Marko Jusup for useful discussions, Xiaoyue Wang, Ju Chen, and Qiwen Zhao for technical help. L.S. was supported by the National Natural Science Foundation of China (Grant No.\ 11931015), Major Program of National Fund of Philosophy and Social Science of China (Grants Nos.\ 22\&ZD158 and ~22VRCO49); C.S. was supproted by JSPS KAKENHI (Grant No. \ JP 23H03499); Z.H. was supported by China Scholarship Council (Grant No. ~202308530309); H.G. was supported by China Postoctoral Science Foundation (Grant No.\ 2023M741852); J.T. was supproted by JSPS KAKENHI (Grants Nos.\ JP 20H02314, JP 23K28189, and ~JP 23H03499); S.H. was supported by Shanghai Artificial Intelligence Laboratory; P.H. was supported by JSPS KAKENHI Grant No.\ JP 21H04595.

\subsection*{Data accessibility} The data and statistical analysis codes used in the study are available at OSF: \href{https://osf.io/s8uzc}{https://osf.io/s8uzc}.

\bibliography{ref}

\clearpage

\onecolumn

\renewcommand\theequation{SI\arabic{equation}}
\renewcommand\thefigure{S\arabic{figure}}
\renewcommand\thepage{S\arabic{page}}
\renewcommand\thetable{S\arabic{table}}
\renewcommand{\thesubsection}{SI\arabic{subsection}}

\setcounter{figure}{1}
\setcounter{page}{1}
\renewcommand{\thepage}{S\arabic{page}}

% \begin{document}

\section*{\centering Supplementary Information for \\
``Beyond a binary theorizing of prosociality''}
% \author{Chen Shen, Zhixue He, Hao Guo, Shuyue Hu, Jun Tanimoto, Lei Shi and Petter Holme}

% \maketitle
% \tableofcontents

% \newpage

\subsection{Supporting Methods}
The experimental framework comprises three parts. In the first part, we designed three treatments to investigate whether zealots promote cooperation in social dilemmas under two different payoff matrices (see Table ~\ref{ts1} for specific values). In each round of the game, participants must simultaneously choose between cooperation and defection. For payoff matrix I (results discussed in the main text), cooperators ($C$) sacrifice one unit for the opponent to receive three units, while defectors ($D$) earn one unit at the expense of one unit from the opponent. For payoff matrix II (results included in the Supporting Information), cooperators give up one unit for the opponent to receive two units, and defectors also earn one unit at the opponent's expense of one unit. These matrices are based on recommendations from reference ~\cite{dreber2008winners}, which pertains to repeated games that allow the same participants to engage multiple times. However, our study is conducted within the framework of a one-shot anonymous prisoner's dilemma game, where participants are never paired with the same opponent twice, ensuring anonymity and randomness in the pairings.

The treatments are structured as follows:

\begin{enumerate}[label={\textbullet}, align=left, leftmargin=2em]
\item Control Treatment: Without Zealots (WZ), where participants are paired with other participants.
\item Scenario 2: Unaware of Zealots (UoZ), where participants are paired with zealots, but are not informed about the existence of zealots.
\item Scenario 3: Informed of Zealots (IoZ), where participants are paired with zealots and are explicitly informed that their opponents are zealots.
\end{enumerate}

Each treatment follows a within-subjects design, meaning that in each session, each participant participates in all treatments in a specific order, with each treatment consisting of 15 rounds. To eliminate order effects, we determined the sequence of treatments in each session using the Latin square method, and each participant engages in each treatment twice per session. This part of the experiment involves six sessions, namely 3 (treatments based on Latin square) × 2 (payoff matrix).

For the second part, we adopt a between-subjects design using payoff matrix I to explore questions regarding the minimum number of zealots required to promote cooperation and whether a bottleneck effect exists in enhancing cooperation. In this part, the variable for different treatments is the probability that participants are paired with zealots. For each fixed encounter probability, we employ a within-subject design where each participant experiences scenario 2 and scenario 3, conducting two sessions according to the Latin square method. The fixed encounter probabilities of 0.25, 0.35, 0.5, and 0.75 are involved in the practical implementation according to the bisection method.

Additionally, we designed a supplementary experiment: Informed of Zealot Bots (IZB), which consists of three sessions. This experiment investigates whether decreasing the motivation to concern for others affects cooperation levels compared to the treatments in the first part. Based on payoff matrix I and payoff matrix II, participants are paired with zealots and are explicitly informed that they are playing with a zealot bot, with only the participants receiving money and the bot not gaining anything from the human side.

In this experimental framework, at the end of each session, each participant fills out a questionnaire about resource allocation to measure their social value orientation. They choose their preferred allocation scheme from the given options, deciding the amount to be allocated to themselves and their opponents.

\subsubsection{Player Recruitment}
The experimental framework was implemented from October 2023 to March 2024. We recruited a total of 964 voluntary undergraduates from Yunnan University of Finance and Economics, Kunming, China, with 52.9\% women, and a mean age of 19.1 years old (Table \ref{ts2}). Each voluntary participant in the student pool was randomly assigned to one of 20 sessions across three parts, ensuring that no subject participated in more than one session of the experiment. Specifically, 244 participants were assigned to the first part of the experiment. For the second part, 480 participants were assigned, with 120 participants in each of the experimental treatments with encounter probabilities of 0.25, 0.35, 0.5, and 0.75, respectively, and each session containing 40 participants. In the third part, 240 participants were assigned, with each session also containing 40 participants.

To mitigate the impact of onymous interaction, upon arrival at the computer lab, participants were randomly assigned to isolated computer cubicles. We ensured anonymity throughout the experiments to maintain the integrity of the results.

\subsubsection{Experimental Implementation}
\begin{figure}[!htbp]
    \centering
    \includegraphics[width=0.9\linewidth]{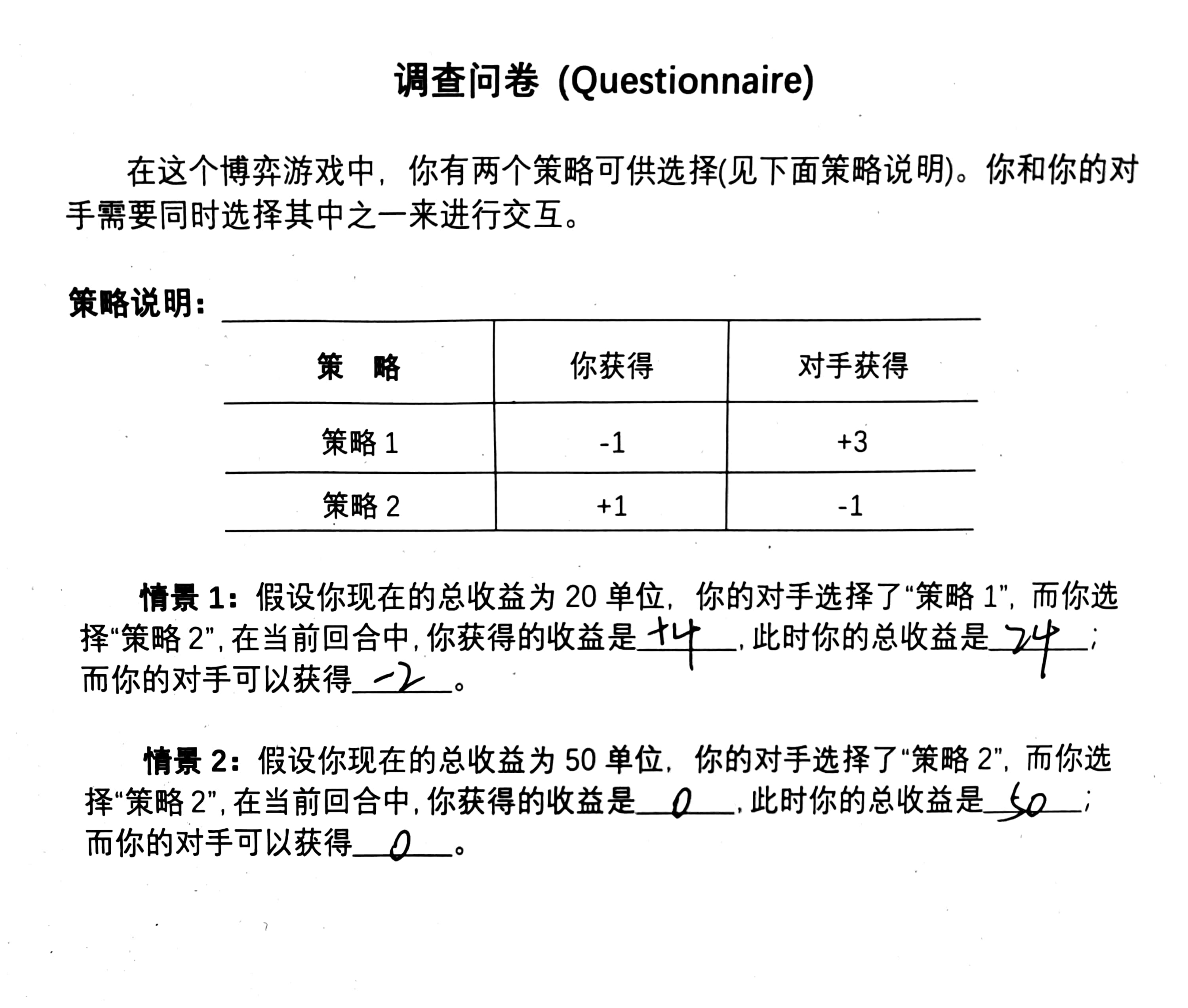}
    \caption{
    \textbf{Snapshot of the questionnaire used to test the basic understanding of PD games.} An English 
translation is as follows. In this game, you have two strategies to choose (see strategy descriptions below). Both you and your opponent need to simultaneously select one of them to engage in interaction. Scenario 1: assuming your current total earnings are 20 units, your opponent chooses ``Strategy 1'' while you choose ``Strategy 2'', in the current round, you earn \_\_\_\_ units and your total earnings become \_\_\_\_ , while your opponent earns \_\_\_\_ units. Scenario 2: assuming your current total earnings are 50 units, your opponent chooses ``Strategy 2'' while you choose ``Strategy 2'', in the current round, you earn \_\_\_\_ units and your total earnings become \_\_\_\_ , while your opponent earns \_\_\_\_ units.
}
        \label{figs1}
\end{figure}

Upon viewing the instructions on their computer screen, each participant completed a simple questionnaire (see Figure~\ref{figs1}) to assess their understanding of the experimental rules. Our staff provided assistance with explanations, and the formal experiment commenced only after all participants had correctly answered the questionnaire. The gameplay interface was developed using the oTree platform~\cite{chen2016otree}.

The total number of rounds varied across the experiment's three parts: the first part comprised six stages with 90 rounds each, the second part had two stages with 30 rounds each, and the third part included a single stage with 15 rounds. In each round, participants had 30 seconds to make a decision through a personalized experiment selection interface, followed by 30 seconds to review the results in the result interface. Since each participant experienced different scenarios in the first and second parts of the experiment, we did not explicitly inform participants about the switch or the differences between scenarios; instead, we allowed them to perceive these changes by observing the experimental interface.

\begin{figure}[!t]
    \centering
    \includegraphics[width=0.84\linewidth]{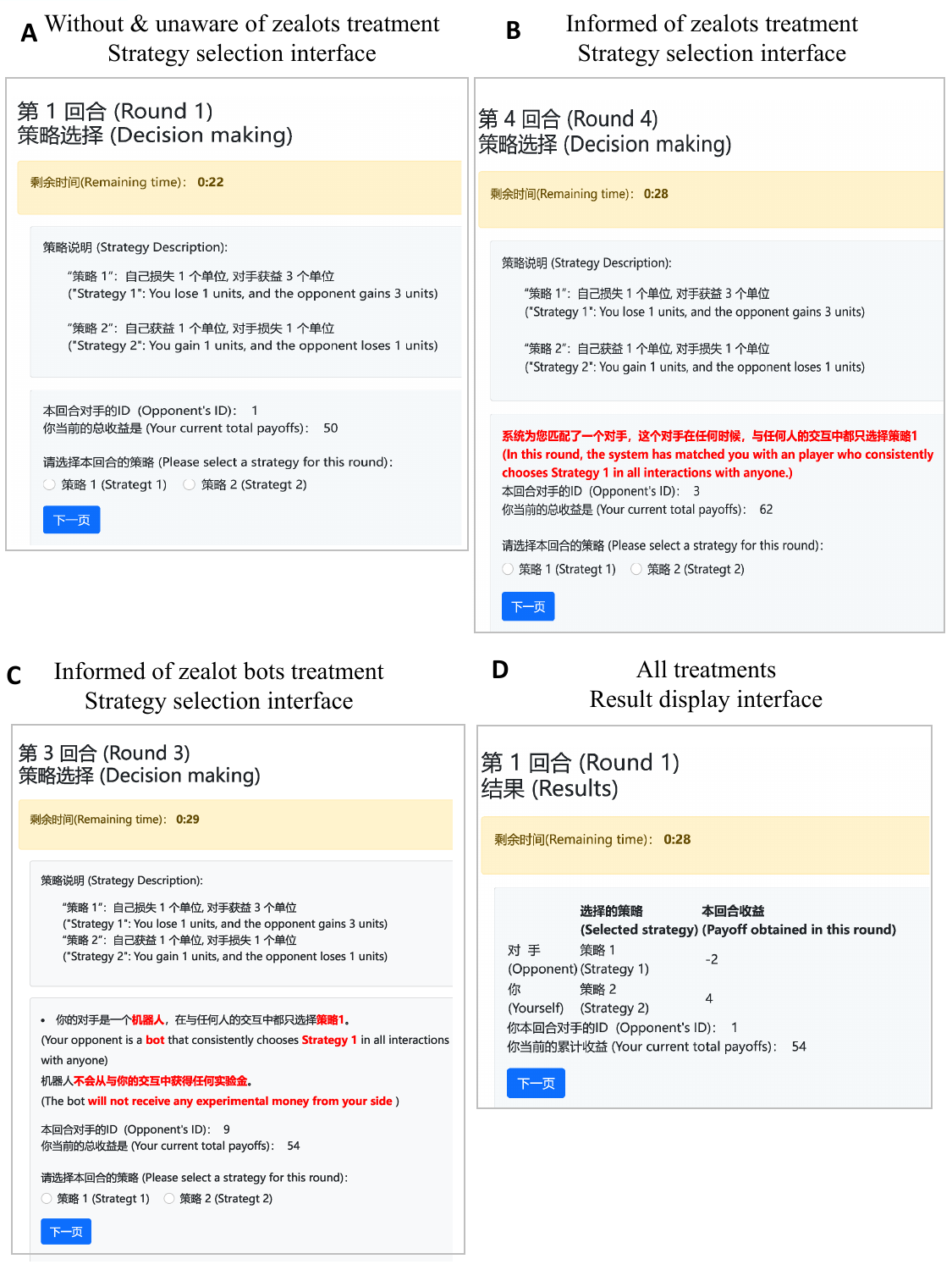}
    \caption{
    \textbf{Interfaces for Different Treatments:} Interface A illustrates the strategy selection interface for participants in treatments without zealots and those unaware of zealots, providing their opponent's ID and the current total payoff for participants. Interface B illustrates the interface for participants encountering zealots in the informed of zealots treatment, with a prompt stating, ``In this round, you are matched with a player who consistently chooses Strategy 1 in all interactions." Interface C shows the interface for the informed of zealot bots treatment, where participants are informed their opponent is a bot that consistently chooses Strategy 1 and cannot receive any monetary gains, highlighted by the prompt, ``Your opponent is a bot that always chooses Strategy 1 and cannot receive experimental money from you." Interface D presents outcome information across all four experimental treatments, detailing the strategies and payoffs for both participants and opponents in the current round, opponent's ID, and participants' cumulative payoffs.}
        \label{figs2}
\end{figure}

The selection interface presented participants with basic information, including their opponent's ID, their current total payoff, and the buttons for making their choice (see Figure \ref{figs2}A for treatment involving without zealots and unaware of zealots). In the Informed of zealots treatment, additional information displayed on the selection interface explicitly informed participants that their opponent is a zealot who always chooses strategy 1 in all interactions with anyone (see Figure \ref{figs2}B). Conversely, in the Informed of zealot bots treatment, participants were further explicitly informed that their opponent is a zealot bot, which cannot receive any experimental money from their side (see Figure \ref{figs2}C). The result interface remained consistent across all experiment treatments, displaying the strategy and payoff for both themselves and their opponent, along with their opponent's ID and their current total payoff (Figure \ref{figs2}D).

\begin{figure}[!t]
    \centering
    \includegraphics[width=0.9\linewidth]{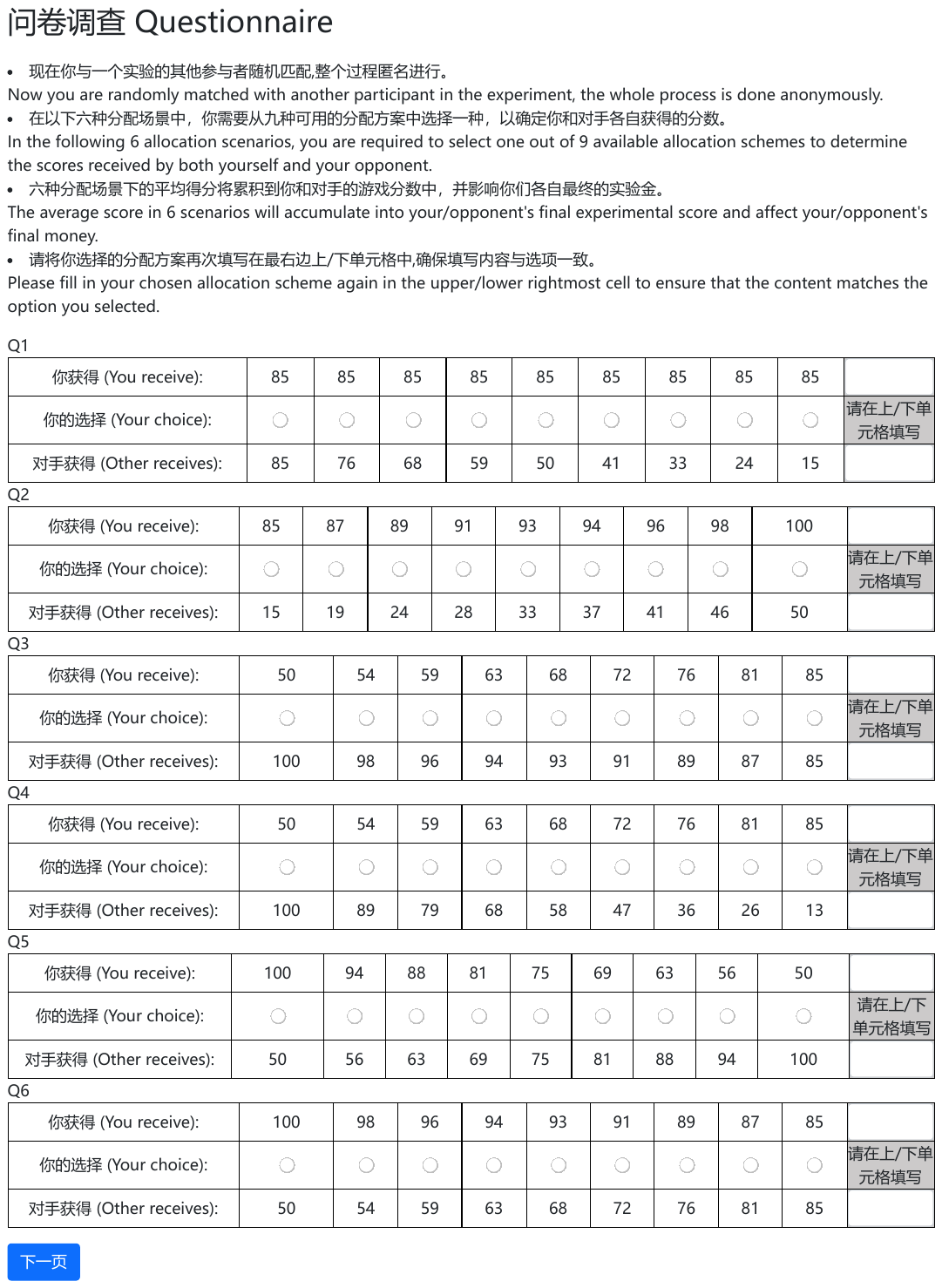}
    \caption{ \textbf{Measuring participants' social value orientations using questionnaires.} Participants are informed that they will be matched with another player.  In the given six scenarios, participants are required to choose a preferred allocation scheme from nine options, where each scheme involves the points that both themselves and their opponent can receive. The average score of the six allocation scenarios directly contributes to the participants' total score, impacting their final monetary payoffs.}
        \label{figs3}
\end{figure}

\begin{figure}[!t]
    \centering
    \includegraphics[width=0.9\linewidth]{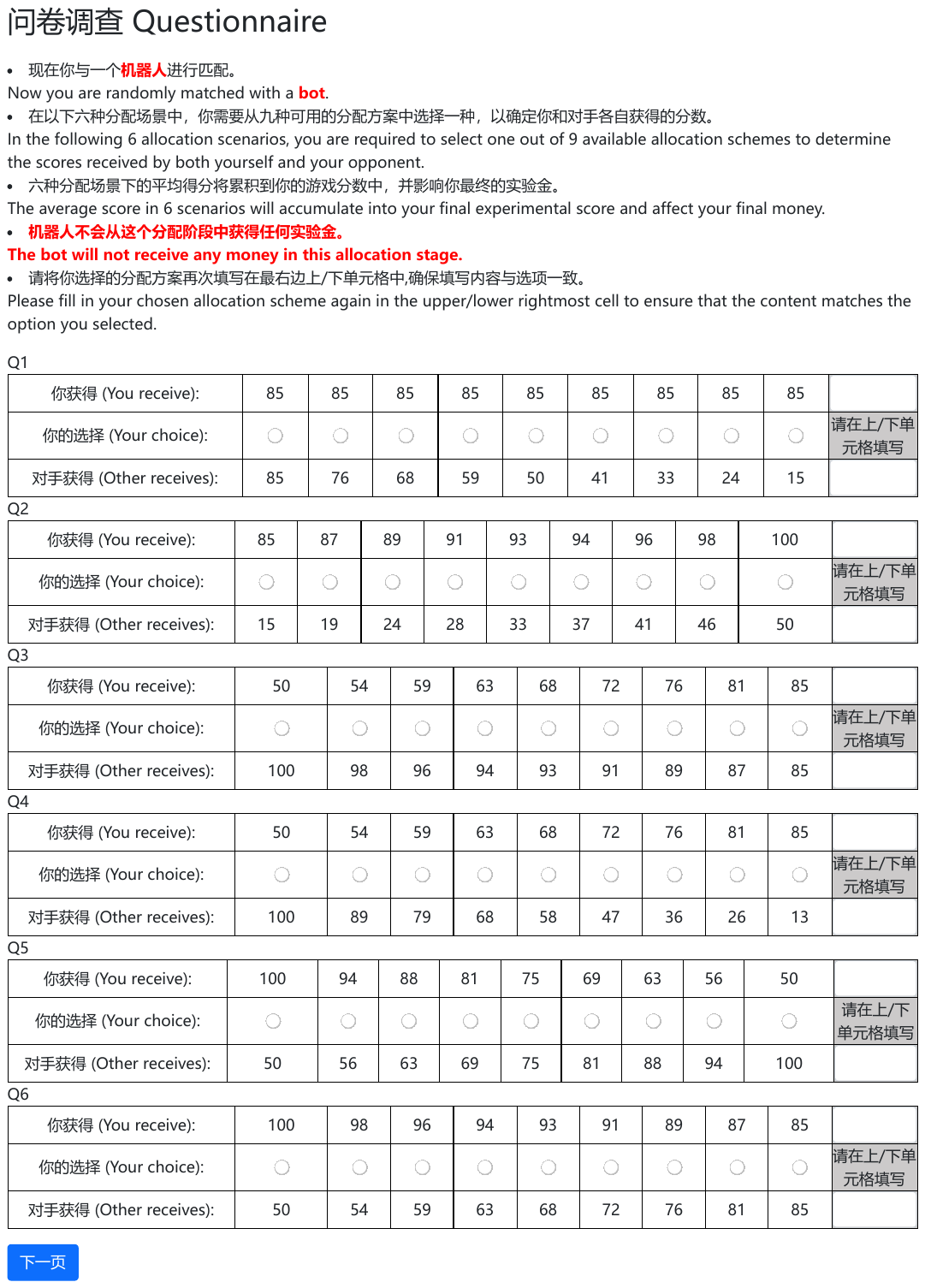}
    \caption{ 
    \textbf{Measuring participants' social value orientations using questionnaires in a design excluding prosocial preferences.} Participants are informed that they will be matched with a bot, and the bot cannot receive any money in this allocation. In the given six scenarios, participants are required to choose a preferred allocation scheme from nine options, where each scheme involves the points that the participants themselves can receive. The average score of the six allocation scenarios directly contributes to the participants' total score, impacting their final monetary payoffs.}
    \label{figs4}
\end{figure}

At the end of each session, participants in Parts 1 and 2 of the experiment completed the questionnaire shown in Figure ~\ref{figs3}, while those in Part 3 filled out the questionnaire in Figure ~\ref{figs4}. In the latter case, participants were further informed that their opponent was a bot and would not receive any money in the allocation phase.

Participants were incentivized by translating their final tokens into a realistic monetary payoff at a rate of 0.5 CNY per token. Additionally, each participant received a 15 CNY show-up fee. If a participant's accumulated tokens were negative, they could not receive any additional benefits beyond the show-up fee. Finally, participants were required to provide their signature to verify their payoffs. On average, participants earned 82.09 CNY, with earnings ranging from 28 CNY to 180 CNY.

Here follows an English translation of gameplay instructions as displayed to volunteers in experimental treatment before the beginning of the game.  We minimally adjusted the text where necessary to suit the needs of experimental control.

\begin{center}
  \begin{minipage}{0.85\textwidth}
\textbf{Instruction}

Welcome to our game experiment!

Please read the following instructions carefully. If you experience any problems during the game, please raise your hand, and our expert staff will help you. This is an anonymous experiment; a computer system will assign everyone a random ID number that cannot be linked back to you. During the game, please refrain from attempting to communicate with other players.

\begin{enumerate}[align=left, leftmargin=1em]
\item Gameplay rules: 
The game consists of a series of rounds where you will be randomly paired with a new opponent in each round. You must choose between `Strategy 1' and `Strategy 2' to interact with your opponent. After making your decision, your payoff will be determined based on the choices made by both you and your opponent. Specifically, if you choose `Strategy 1', you will receive $-1$ token, and your opponent will receive $+3$ tokens. If you choose `Strategy 2', you will receive $+1$ token, and your opponent will receive $+3$ tokens. The outcomes for mutual choices are as follows: if both choose `Strategy 1', each of you will receive 2 units; if both choose `Strategy 2,' each will receive 2 units. However, if you choose `Strategy 1' and your opponent chooses `Strategy 2', you will receive -2 units while your opponent receives +4 units. Conversely, if you choose `Strategy 2' and your opponent chooses `Strategy 1,' you will receive +4 units, and your opponent will receive -2 units. Below is a summary of the payoff calculations:

\begin{equation*}
\begin{array}{cccc}
\hline
\text{Option} & \text{You} & \text{Opponent} \\
\hline
\text{Strategy 1} & -1 & +3 \\
\text{Strategy 2} & +1 & -1 \\
\hline
\end{array}\Longrightarrow
\begin{array}{cccc}
\hline
\text{You/Opponent} & \text{Strategy 1} & \text{Strategy 2} \\
\hline
\text{Strategy 1} & 2 & -2 \\
\text{Strategy 2} & +4 & 0 \\
\hline
\end{array}
\end{equation*}

\item Experiment interface: Gameplay happens via a custom computer interface consisting of two screens: 

\begin{enumerate}[label={\textbullet}, align=left, leftmargin=2em]
\item On the selection interface, you'll find a description of how each strategy affects your and your opponent's payoff, along with your current total payoff. Additionally, information about your paired opponent, such as their ID number, will be displayed. You have 30 seconds to make your decision. After choosing, click the ``Next" button to proceed to the result screen. Failure to do so within the allotted time will automatically advance the system to the next screen.
\item On the result interface, you'll have 30 seconds to review key information, including your opponent's strategy and payoff, your own strategy and payoff for the current round, and your overall total payoff. Click the ``Next" button to move to the next screen and start a new round. Not pressing the button in time will trigger an automatic transition to the next screen.
\end{enumerate}
\item Questionnaire: At the end of the experiment, you'll need to complete a questionnaire, which will be available in the questionnaire section after the formal experiment begins.
\item  Monetary payout: Upon conclusion of the experiment, you'll see your final accumulated tokens. Our staff will convert these tokens into a real monetary payout at a rate of 0.5 CNY per token. Additionally, you'll receive a show-up fee of 15 CNY regardless of your performance during the experiment.
 \end{enumerate}
  \end{minipage}
\end{center}

\subsection{Supporting results}
\subsubsection{Order effect}
The position of each treatment in the experimental schedule that encompasses all sessions in a collective manner has significant impact on cooperation level for the scenarios of without zealots and unaware of zealots, but not for aware of zealots (see table \ref{ts4}). Since each session was composed by the same participants, and different sessions encompass different participants, and each participants play each treatment twice within each session. The obtained data have both dependent and independent feature, which lead us to adopt the liner mixed model to check the order effect for each treatment. By setting the order as a categorical factor and designating order 1 (the first appearance in the experimental schedule in all sessions) as the reference, we found that when the without zealots treatment is placed third, fourth, or sixth (order 3, 4, and 6 respectively), the cooperation level is significantly lower than those in order 1. Similar order effects were observed in the unaware of zealots treatment, where the cooperation level in order 6 is significantly lower than those in order 1. In contrast, for the aware of zealots treatment, we did not find any order effect; regardless of their position in the experimental schedule, the cooperation level remained similar to those in order 1.

\begin{figure}[!t]
    \centering
    {\includegraphics[width=0.89\linewidth]{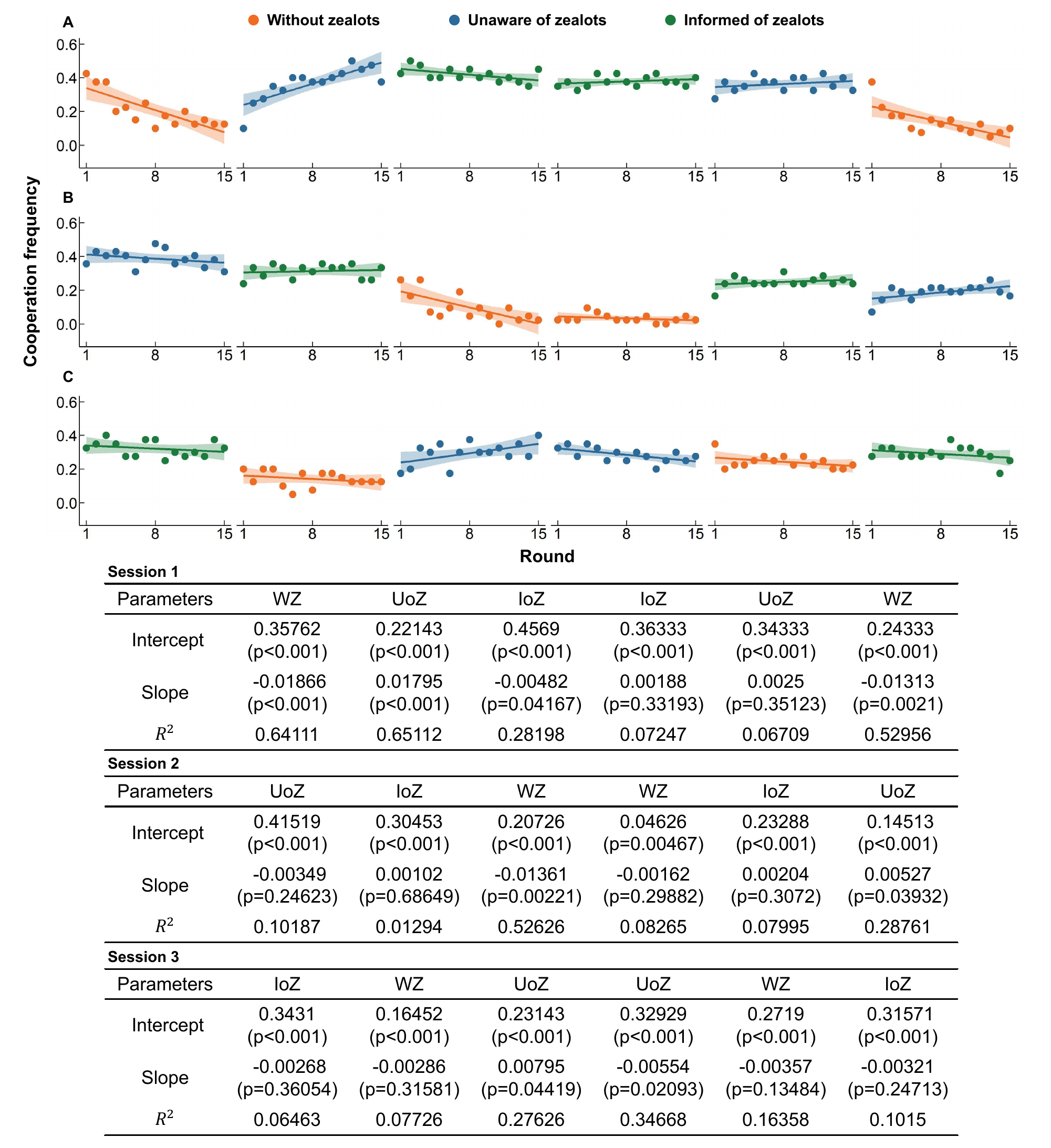}}
    \caption{\textbf{This graph represents the frequency of cooperative choices over 90 successive rounds, divided into 15-round segments for each treatment, in three sessions of a one-shot anonymous prisoner's dilemma game under payoff matrix I.} The treatments include: without zealots (WZ), unaware of zealots (UoZ), and informed of zealots (IoZ). Panels A, B, and C correspond to the three sessions, respectively, showing the mean cooperation frequency per round. The trend lines in each panel reflect the change in cooperation over the 15-round treatments. The accompanying statistical summary for each session lists the intercept, slope, and $R^{2}$, with $p$-values indicating the significance of these parameters.}
        \label{figs5}
\end{figure}

To further explore how cooperation changes over time under each treatment, Figure \ref{figs5} illustrates the time-dependent frequency of cooperative decisions in a within-subjects experiment conducted under payoff matrix I, across three sessions featuring treatments without zealots, unaware of zealots, and informed of zealots. Generally, cooperation is substantially higher when participants engage in treatments involving zealots, as opposed to those without. An exception is noted in the third session's second-to-last treatment without zealots, where the mean cooperation level (24.3\%) is similar to that in the subsequent treatment with participants informed of zealots (mean of 29\%, pairwise $t$-test, $t=-1.6$, $df=39$, $p$-value=0.12). It also aligns closely with the preceding treatments where participants were unaware of zealots (mean of 28.5\%, pairwise $t$-test, $t=-1.35$, $df=39$, $p$-value=0.18). These patterns indicate that participants may either be confused about the rules of the game or may have developed optimistic beliefs about their opponents' propensity to cooperate, despite the absence of zealots.

\subsubsection{Robustness check: Different payoff matrix}
\begin{figure}[]
    \centering
    {\includegraphics[width=0.99\linewidth]{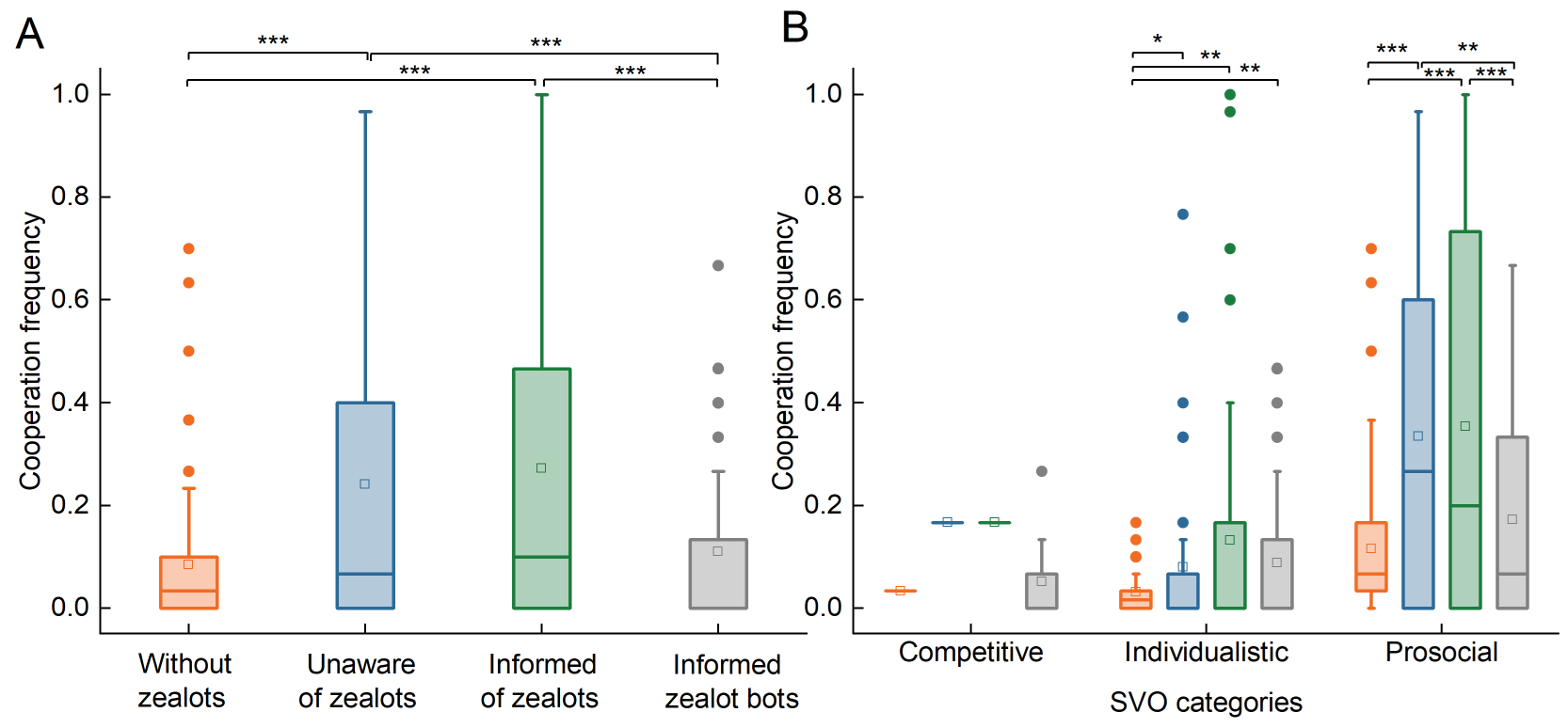}}
    \caption{\textbf{Robustness Check Using Payoff Matrix II.} A: Boxplots show the empirical distribution of cooperation frequencies across different treatments, defined as the ratio of cooperative actions to total rounds per participant. The average cooperation frequency is 8.4\% in the without zealots treatment, significantly lower than 24.2\% in the unaware of zealots treatment (pairwise $t$-test; $t=-6.73$, $p < 10^{-10}$) and 27.2\% in the informed of zealots treatment (pairwise $t$-test; $t=-6.71$, $p < 10^{-10}$). No significant difference exists between the unaware and informed of zealots treatments (pairwise $t$-test; $t=-1.59$, $p$ = 0.11). However, the informed of zealot bots treatment shows a lower average cooperation frequency of 11\%, significantly different from both involving zealot treatments (two-sample $t$-test; $t=4.11$, $p < 10^{-5}$; $t=4.69$, $p < 10^{-6}$) but not from the without zealots treatment (two-sample $t$-test; $t=-1.42$, $p$ = 0.16). B: Boxplots show cooperation frequency distributions within competitive, individualistic, and prosocial categories across treatments. Individualistic participants exhibited significantly higher cooperation in unaware (8\%) and informed (13\%) of zealot treatments compared to the without zealots treatment (3\%; pairwise $t$ tests; $t=-2.09$, $p$=0.04 and $t=-2.8$, $p$=0.008, respectively), with no significant differences in the informed of zealot bots treatment (8\%; two-sample $t$ tests; $t=-0.29$, $p$=0.77 and $t=1.08$, $p$=0.28). The informed of zealot bots treatment showed significantly higher cooperation compared to the without zealots treatment (two-sample $t$-test; $t=-3.34$, $p$=0.001). Prosocial participants exhibited significantly higher cooperation frequencies in the unaware (33.6\%) and informed (35.4\%) of zealots treatments compared to the without zealots treatment (11.6\%; pairwise $t$ test, $t=-6.75$, $p < 10^{-9}$; $t=-6.22$, $p < 10^{-8}$) and the informed of zealot bots treatment (17.2\%; two-sample $t$-test, $t=-3.2$, $p=0.002$; $t=3.39$, $p=0.001$), with no significant difference between the informed of zealot bots and without zealots treatments (two-sample $t$-test, $t=-1.57$, $p=0.12$).
} \label{figs6}
\end{figure}

\paragraph{Overall cooperation level and the role of social value orientation.}
The trend of overall cooperation level across treatments obtained under payoff matrix II mirror those obtained under payoff matrix I. As shown in Figure \ref{figs6}A, cooperation levels significantly increase in treatments involving zealots compared to the without zealots treatment, regardless of whether participants are informed about their opponents' zealotry. However, no significant difference in cooperation frequencies can be observed between the unaware of zealots treatment and the aware of zealots treatment. Notably, a significant decrease in cooperation frequency was observed in the informed of zealot bots treatment, aligning closely with the control treatment.

Figure \ref{figs6}B illustrates cooperation frequencies among participants with different SVO across treatments (refer also to Table~\ref{ts10} for classification results). It is observed that individualistic participants exhibit significantly higher cooperation levels in treatments involving zealots and zealot bots, regardless of their awareness of zealots or the non-materialistic nature of zealot bots, compared to treatments without zealots. Moreover, no significant difference in cooperation level was found between treatments involving zealot bots and those involving zealots, or between treatments involving zealot bots and treatments without zealots. In contrast, the trend of cooperation level for prosocial participants across treatments mirrors the overall cooperation tendency observed when participants with distinct SVO are pooled together.

\begin{figure}[]
    \centering
    {\includegraphics[width=0.81\linewidth]{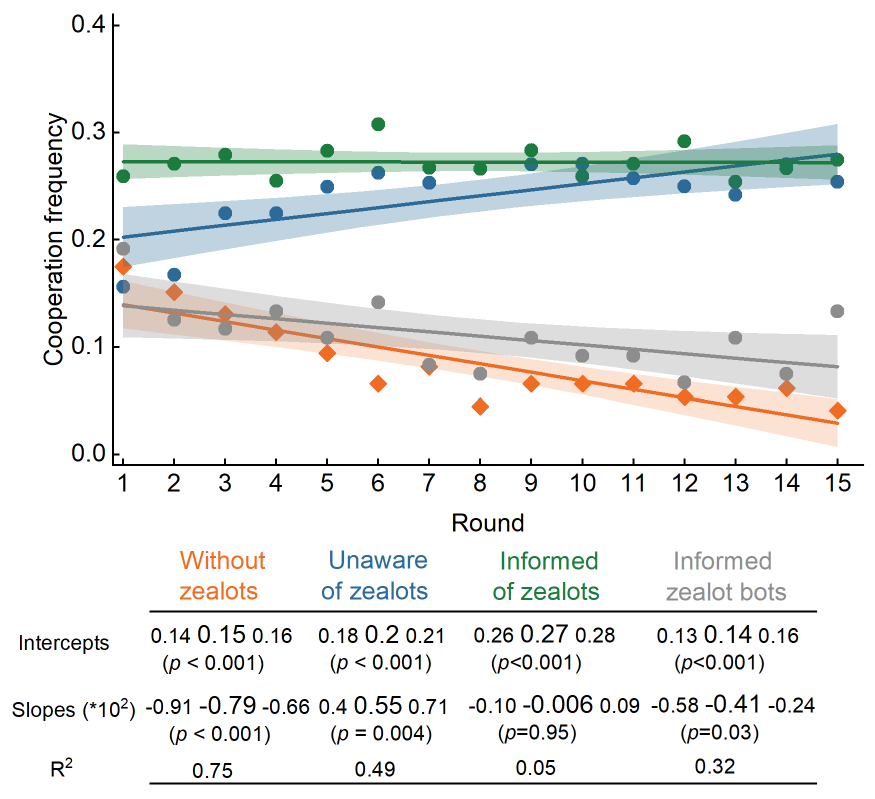}}
    \caption{\textbf{The presence of zealots stabilises cooperation.} Shown are the frequency of cooperation over time, obtained under payoff matrix II, illustrated by data points and regression lines for three distinct treatments: without zealots, unaware of zealots, and informed of zealots. In the treatment without zealots, the frequency of cooperation exhibits a slowly decreasing trend, with a slope significantly less than zero. Conversely, in treatments involving zealots—whether participants were unaware of the zealots or informed — the frequency of cooperation either shows a slowly increasing trend or remains stable, indicating that the presence of zealots has a stabilizing impact on cooperation levels. The frequencies were calculated by dividing the number of volunteers who chose to cooperate by the total number of participants in each round. Smaller font sizes and shaded areas represent 95\% confidence intervals.}
        \label{figs7}
\end{figure}

\paragraph{Time series analysis of Players' cooperation frequency.}
The cooperation trend over time within each treatment obtained under payoff matrix II generally mirrors those obtained under payoff matrix I, with one exception where cooperation significantly decreases in the informed of zealot bots treatment. Figure~\ref{figs7} displays the frequency of cooperation over time, aggregated across three sessions. In the treatment without zealots, cooperation levels start at 17.5\% and significantly decrease to 4\% by the final round. Conversely, in the unaware of zealots treatment, cooperation levels start at 15.6\% and significantly increase to 25.4\% by the final round. In the informed of zealots treatment, cooperation levels start at 25.8\% and remain stable over time, with the final round showing a cooperation level of 27.5\%. However, in the informed of zealot bots treatment, cooperation levels start at 19.2\% and significantly decrease to 13.3\% by the final round.

Comparing to the treatment without zealots, the initial cooperation levels in the treatments involving unaware of zealots and informed of zealot bots are not significantly different. However, the initial cooperation level in the informed of zealots treatment is significantly higher than that in the treatment without zealots. Additionally, the final cooperation levels in the treatments of unaware of zealots, informed of zealots, and informed of zealot bots are all significantly higher than that in the treatment without zealots. These results are summarized in Table~\ref{ts11}.

\paragraph{Order effect.}
Similar to the results obtained under payoff matrix I, we also observe that the position of each treatment in the experimental schedule under payoff matrix II significantly impacts the cooperation level in scenarios without zealots and unaware of zealots, but not in informed of zealots scenarios. Setting order 1 as the reference, we find that when the without zealots treatment is placed fourth and sixth, the cooperation level significantly decreases. In the unaware of zealots treatment, the cooperation level in order 6 is significantly lower than in order 1. However, for the aware of zealots treatment, we did not find any order effect; regardless of their position in the experimental schedule (see Table~\ref{ts12} for details).

\begin{figure}[!t]
    \centering
    {\includegraphics[width=0.91\linewidth]{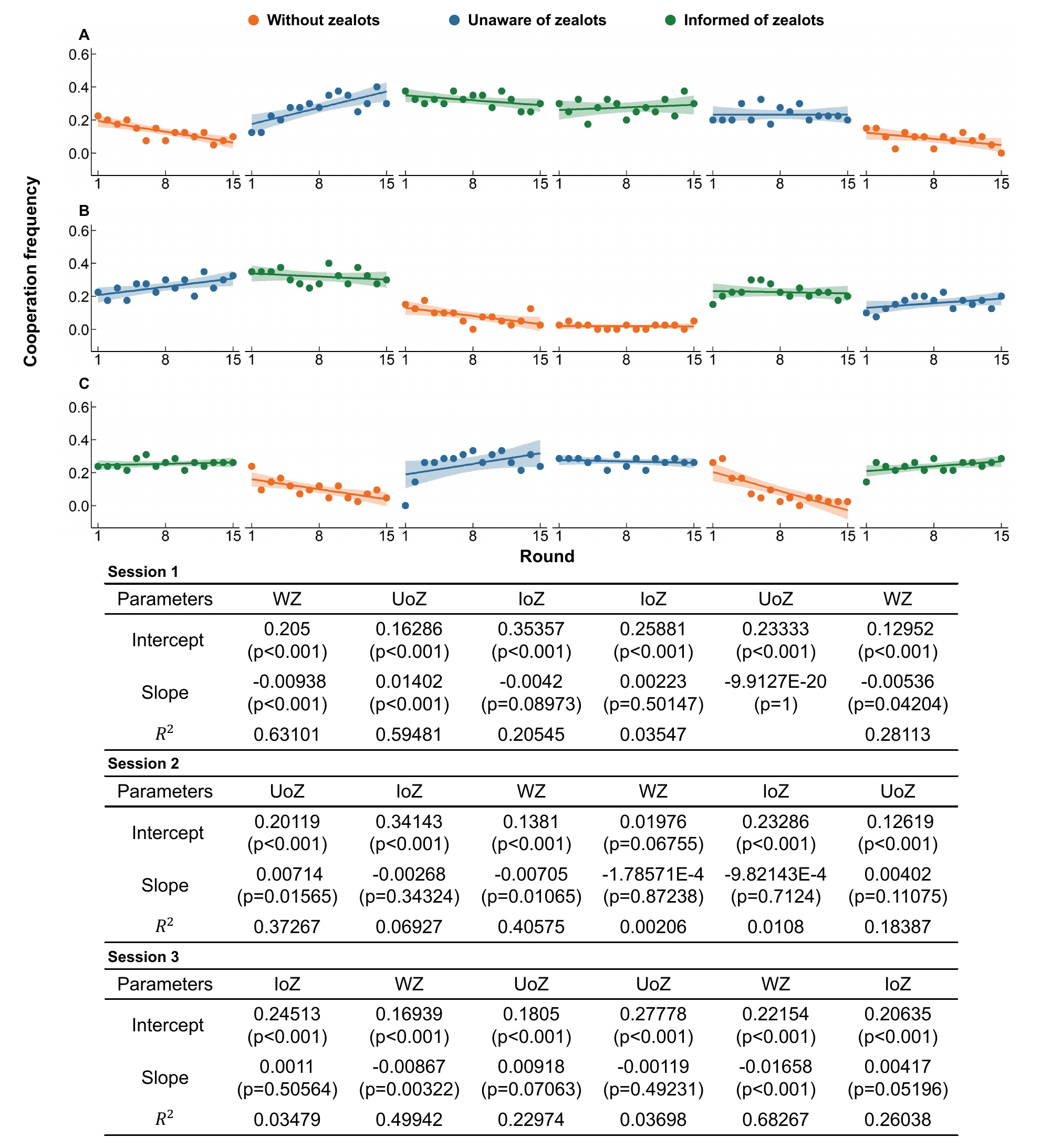}}
    \caption{\textbf{The figure presents the frequency of cooperative decisions across three sessions in a one-shot anonymous prisoner's dilemma game, each session consisting of 90 rounds segmented into 15-round treatments, under payoff matrix II.} The three treatment conditions examined are: without zealots (WZ), unaware of zealots (UoZ), and informed of zealots (IoZ). Panels A, B, and C represent the sessions, showing the mean cooperation frequency per round with trendlines. The table below each graph provides statistical metrics such as intercept, slope, and $R^{2}$ values, including $p$-values for significance testing. }
        \label{figs8}
\end{figure}

The trends of cooperation over time under each treatment for each session under payoff matrix II are consistent with those obtained under payoff matrix I. However, we did not observe the exception where cooperation levels in treatments involving zealots were not higher than those in treatments without zealots across sessions (see Figure~\ref{figs8}).

\paragraph{Payoff distribution among players with different SVO across treatments under payoff matrix II.} Comparative analysis of payoffs obtained in the context of Payoff Matrix II yields consistent conclusions with those obtained under Payoff Matrix I. Prosocial individuals demonstrate a willingness to sacrifice part of their interest for the collective good, yet they also exhibit self-interest as their payoffs consistently exceed those of their zealot opponents.

\begin{figure}[!t]
    \centering
    {\includegraphics[width=0.88\linewidth]{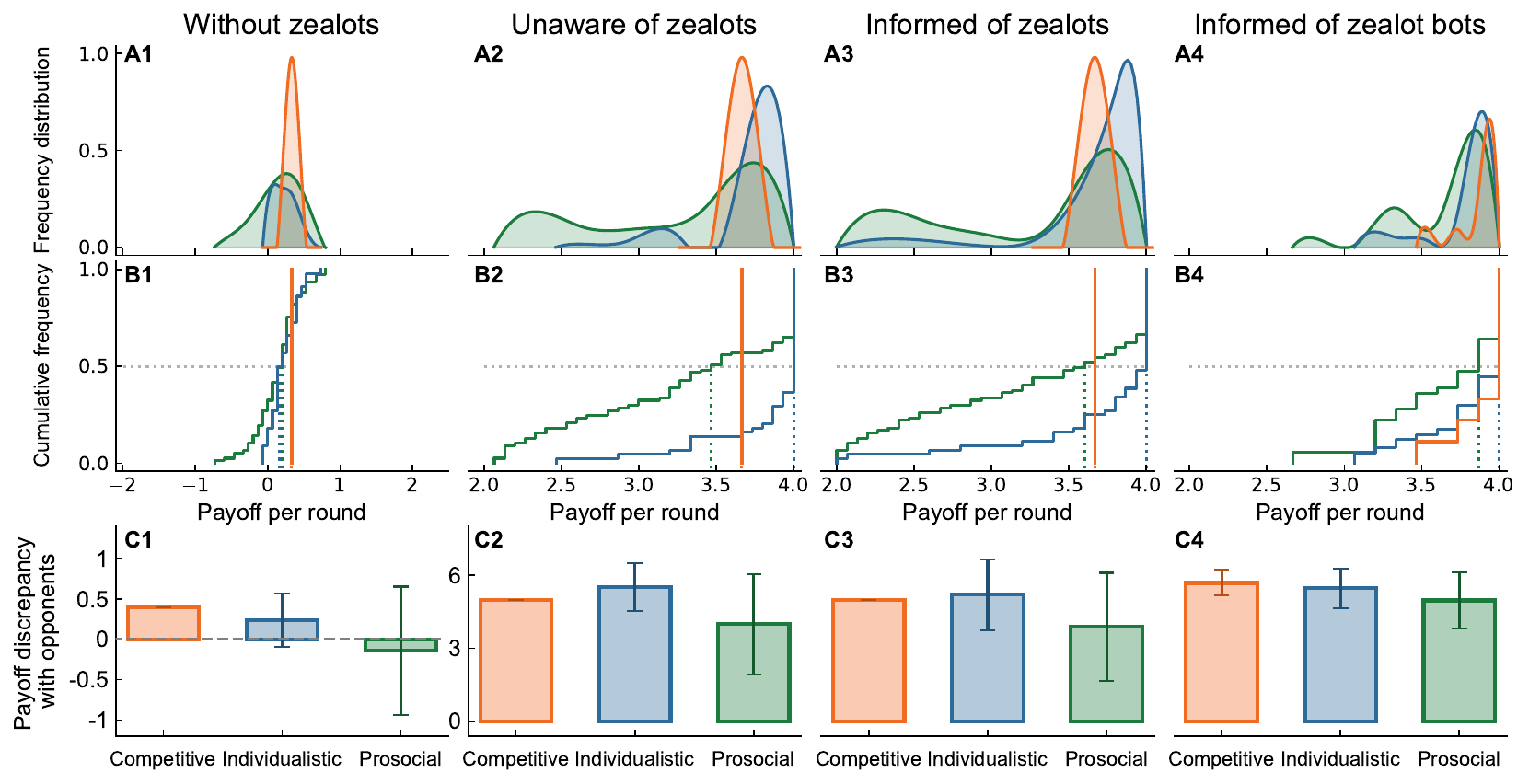}}
    \caption{\textbf{Analysis of payoffs among participants with different SVO types across treatments, obtained using Payoff Matrix II}. \textbf{Panel A} presents the distribution of per-round payoffs among participants identified as competitive, individualistic, and prosocial, for each treatment: without zealots, unaware of zealots, informed of zealots, and with informed of zealot bots, displayed sequentially from left to right. \textbf{Panel B} presents the cumulative density differences in payoffs among SVO categories. For treatment without zealots, prosocial players (median value 0.2) do not significantly differ from individualistic players (median values 0.167, Mann–Whitney $U$ test: $U = 1423.5, p = 0.144$). However, with zealots introduced, prosocial players consistently achieve lower median payoffs compared to individualistic players across all treatments (median values 3.467, 3.6, and 3.867; Mann–Whitney $U$ test: $U = 977.5, p < 0.001; U = 1143.5, p = 0.002; U = 1000, p = 0.024$ in the unaware of zealots, informed of zealots, and informed of zealot bots treatments, respectively).  \textbf{Panel C} provides bar charts illustrating the payoff difference between players and opponents among participants with distinct SVO types across treatments. Individualistic players consistently gain higher payoffs than their opponents under all treatments (from panel C1-C4, mean values of payoff difference: 0.236, 5.52, 5.2, and 5.47; one sample $t$-test, $t = 6.05, p < 0.001; t = 49.03, p < 0.001; t = 31.45, p < 0.001; t = 76.57, p < 0.001$). Prosocial players also exhibit significant differences compared to their opponents (from panel C1-C4, mean values of payoff difference: -0.14, 3.99, 3.88, and 4.97; one sample $t$-test,$ t = -2.02, p = 0.045; t = 22.79, p < 0.001; t = 20.39, p < 0.001; t = 34.10, p < 0.001$). The payoff differentials of prosocial players are significantly lower than those of individualistic players (median values: 0.0 and 0.2, Mann–Whitney $U$ test: $U = 1148.5, p = 0.003$ for without zealots treatment; median values: 4.4 and 6.0, $U = 977.5, p < 0.001$ for unaware of zealots treatment; median values: 4.8 and 6.0,  
    $U = 1143.5, p = 0.002$ for informed of zealots treatment; median values: 5.6 and 6.0, $U = 1000, p = 0.024$ for informed zealot bots treatments). Across treatments, there is no significant alteration in the payoff differentials for prosocial individuals (median values: 4.4, 4.8, and 5.6, Kruskal-Wallis test: $H=3.977, p = 0.137$), and the same holds true for individualistic individuals (median values: 6.0, 6.0, and 6.0, Kruskal-Wallis test: $H=1.514, p = 0.469$).
 }
        \label{figs9}
\end{figure}

Figure~\ref{figs9}A-C displays the distributions of payoffs per round among participants with distinct SVO types across treatments, the accompanying cumulative density among participants with distinct SVO types across treatments, and the bar charts of the payoff difference with opponents among participants with distinct SVO types across treatments. Due to the limited number of competitive participants in treatments without and involving zealots (see also Table~\ref{ts10}), we skip the analysis of competitive participants and focus on prosocial and individualistic players.

In the treatment without zealots, the distributions of payoffs per round between individualistic and prosocial individuals are similar (Figure~\ref{figs9}A1), with no significant differences in median payoffs (Figure~\ref{figs9}B1). In contrast, introducing zealots significantly shifts this result, with prosocial players earning less than their individualistic counterparts, regardless of whether they are aware of the zealots (Figure~\ref{figs9}A2-A3 and B2-B3). Although disclosing the non-materialistic nature of zealot bots to participants increases the average payoff of prosocial participants, it remains significantly lower than that of individualistic individuals (Figure~\ref{figs9}A4 and B4).

Lastly, examining the difference in payoffs between players and their opponents reveals that prosocial individuals lean towards self-interest. In treatments involving zealots or zealot bots, we observe that both individualistic and prosocial players consistently gain significantly more than their zealot opponents, irrespective of whether they are aware of the zealots' presence or the non-materialistic nature of zealot bots (see Figure~\ref{figs9}C2-C4).

\subsubsection{The impact of awareness of zealots on cooperation across treatments varying encounter probabilities} In scenarios where the probability of encountering zealots is less than 1, participants have the opportunity to interact with both zealots and other participants. Consequently, in the informed of zealots treatment, we distinctively measure cooperation frequencies for rounds where participants are unaware they are playing against regular participants and for rounds where they are informed of opposing zealots. Conversely, the unaware of zealots treatment uniformly lacks opponent information, regardless of whether they are zealots. We examine the effect of disclosing opponents' zealotry on participants' cooperation levels across treatments with varying encounter probabilities: 25\%, 35\%, 50\%, and 75\%.

\begin{figure}[!t]
    \centering
    {\includegraphics[width=0.91\linewidth]{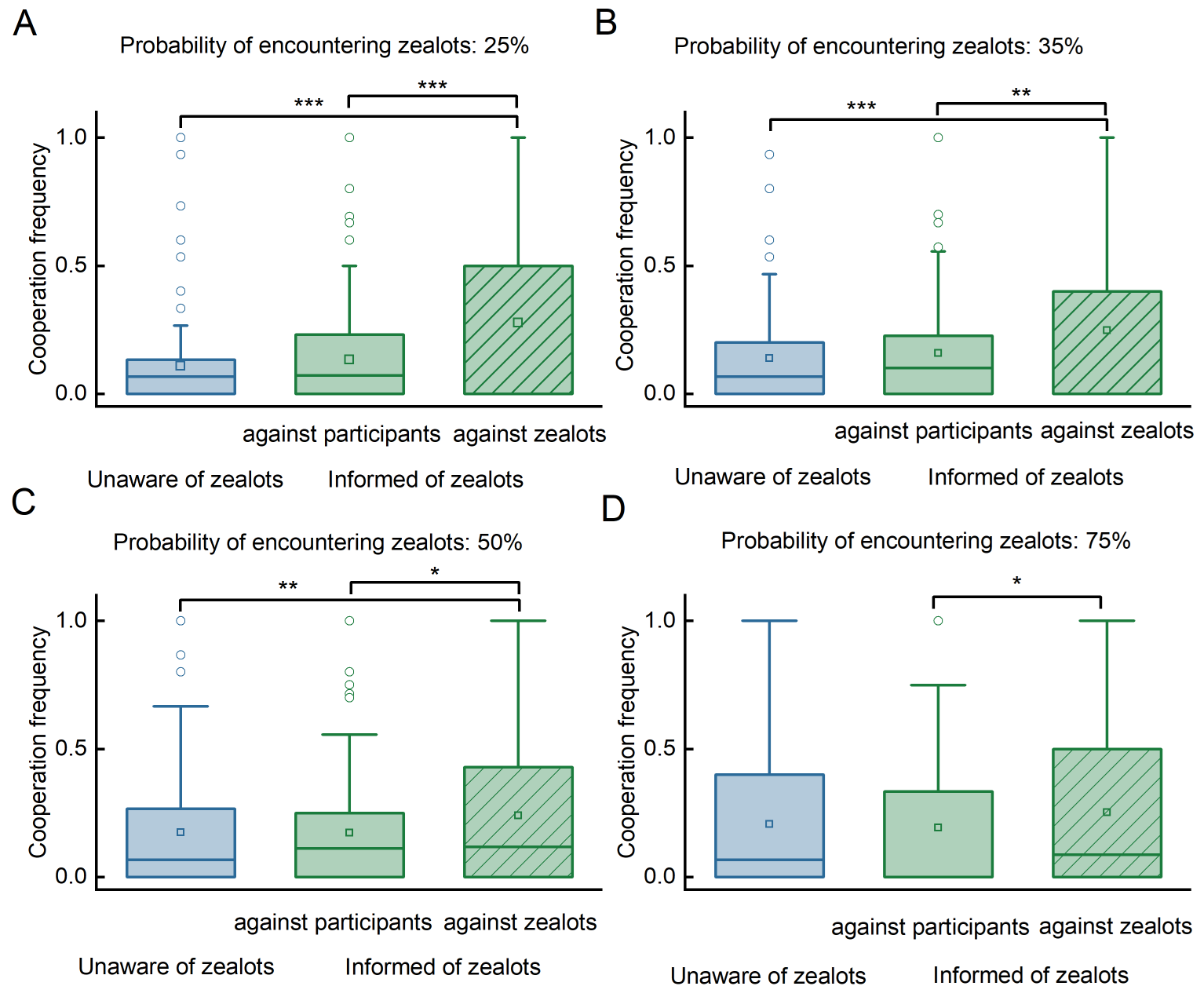}}
    \caption{\textbf{The effect of participant awareness of zealot on cooperation levels at various encounter probabilities ($q$).} Panels A-D illustrate the effect of being informed about zealots on participants' cooperation levels, with comparison to when no information is provided (unaware of zealots treatment), across varying probabilities of encountering zealots: 25\%, 35\%, 50\%, and 75\%.  In the informed of zealots treatment, participants’ cooperation frequency is separately recorded for rounds against other participants (where they are not informed) and against zealots (where they are informed). The unaware of zealots treatment consistently omits information about opponents, whether they are zealots or not. Notably, in the informed of zealots treatment, participants show higher cooperation levels when informed about their opponents' zealotry as opposed to when such information is withheld. This significant effect persists across all probabilities of zealot encounters (Paired sample $t$ test; mean value 13.3\% vs. 27.6\%; $t=-4.1$ $p<0.0001$; mean value 13.3\% vs. 27.6\%; $t=-4.1$ $p<0.0001$ for panel A; mean value 13.3\% vs. 27.6\%; $t=-4.1$ $p<0.0001$; mean value 15.8\% vs. 24.7\%; $t=-2.98$ $p<0.004$ for panel B; mean value 17.2\% vs. 24.0\%; $t=-2.37$ $p=0.02$ for panel C; mean value 19.1\% vs. 25.2\%; $t=-2.36$ $p=0.02$ for panel D). Additionally, informed participants exhibit significantly higher cooperation than those in the unaware of zealots treatment, with the exception at a 75\% encounter probability (Paired sample $t$ test; mean value 11\% vs. 27.6\%; $t=5.12$ $p<0.0001$ for panel A; mean value 13.8\% vs. 24.7\%; $t=3.64$ $p<10^{-4}$ for panel B; mean value 17.3\% vs. 24.0\%; $t=-2.70$ $p=0.007$ for panel C; mean value 20.7\% vs. 25.2\%; $t=-1.83$ $p=0.07$ for panel D). Vertical bars denote standard errors. Significance levels are noted: *$p<0.05$, **$p<0.01$, ***$p<0.001$, and non-significant differences are not marked with an asterisk.
    }
        \label{figs10}
\end{figure}

As depicted in Figure \ref{figs10}, participants in the informed of zealots treatment consistently demonstrate higher cooperation levels when they are aware of the zealotry of their opponents compared to when they are uninformed. This trend is significant at all tested probabilities of encountering zealots. Furthermore, participants informed about zealot encounters exhibit significantly higher cooperation levels than those who remain unaware, except at the 75\% probability threshold. In addition, the distribution of SVO categories across these treatments does not significantly differ (refer to Table \ref{ts13}).

\subsubsection{Gender as a confounding factor}
Gender could be a confounding variable impacting our experiment's results. To explore its influence, we used contingency table tests (refer to Table \ref{tab_gender}). Statistical analyses indicate that at a zealot encounter probability $q$ of 1, females show significantly lower levels of cooperation than males in the without zealots, unaware of zealots, and informed of zealots treatments. Interestingly, this gender disparity does not persist under payoff matrix II, where cooperation levels between genders are statistically indistinguishable. However, in the informed zealot bots treatment, female cooperation is notably less than male cooperation. At varying levels of $q$, except for the informed of zealots treatment at $q=0.5$, where gender does not seem to influence cooperation, females consistently exhibit lower cooperation rates than males.

\subsubsection{Academic background as a confounding factor}
The academic disciplines of students participating in our experiment, categorized as either mathematics and natural sciences (M \& NS) or humanities and social sciences (H \& SS), could influence the outcomes. To assess the impact of academic background, we conducted contingency table tests (see Table \ref{tab_major}). Analysis shows that at a zealot encounter probability $q$ of 1 in the informed of zealots treatment, M \& NS students display significantly lower cooperation than H \& SS students. In contrast, no discernible difference in cooperation levels between the disciplines is evident when participants interact with bots. As $q$ varies, except at $q=0.5$ where M \& NS students cooperate significantly less than H \& SS students, cooperation levels across other treatment conditions show no substantial differences related to students’ fields of study.

\clearpage

\subsection{Supporting Figures}

\clearpage

\clearpage

\clearpage

\clearpage

\clearpage

\clearpage

\clearpage
\clearpage

\clearpage

\clearpage
%\begin{figure}[!htbp]
%    \centering
%    {\includegraphics[width=0.99\linewidth]{SI_figure/fig_payoff_prob.pdf}}
%    \caption{
%    \textbf{The per round payoff of player under different probabilities of encountering zealots.} In the treatments of ``unaware zealots'' and ``informed of zealots,'' players in all four encounter probability conditions exhibit significantly higher payoff than their opponents. Under the treatment of encountering ``unaware zealots,'' in the encounter probabilities of 0.35 and 0.5 conditions, the prosocial players'  payoff are significantly lower than those of individualistic players (Mann–Whitney $U$ test: $U=2012$, $p=0.012$; $U=2131$, $p=0.004$, respectively). Among the other encounter probability conditions, there is no significant difference in earnings among the three player types. Under the treatment of ``informed of zealots,'' in the encounter probabilities of 0.35 and 0.75 conditions, the prosocial players' are significantly lower than those of individualistic players (Mann–Whitney $U$ test: $U=1034$, $p=0.001$; $U=1210$, $p=0.008$ respectively). Among the other encounter probability conditions, there is no significant difference in payoff among the three player types.}
%        \label{fig:payoff_prob}
%\end{figure}

\subsection{Supporting Tables}

\begin{table}[htbp]
  \caption{The two payoff matrices used in the experiment. Each cell indicates the payoff for the row player (Player A) followed by the payoff for the column player (Player B). Defection is the strict Nash equilibrium in both payoff matrices.}
    \centering
\begin{minipage}[t]{0.4\linewidth}
\resizebox{1\linewidth}{!}{
\begin{tabular}{cccc}
\multicolumn{4}{c}{Payoff matrix I} \\ 
\hline
   &  & \multicolumn{2}{c}{Player B}  \\
   &  & Strategy 1 ($C$) & Strategy 2 ($D$) \\
   \hline
  \multirow{2}{*}{Player A} & Strategy 1 ($C$) & $2,2$ & $-2,4$ \\
  & Strategy 2 ($D$) & $4,-2$ & $0, 0$ \\
  \hline
\end{tabular}
}
\end{minipage}
\begin{minipage}[t]{0.4\linewidth}
\resizebox{1\linewidth}{!}{
\begin{tabular}{cccc}
\multicolumn{4}{c}{Payoff matrix II} \\ 
\hline
   &  & \multicolumn{2}{c}{Player B}  \\
   &  & Strategy 1 ($C$) & Strategy 2 ($D$)\\
   \hline
  \multirow{2}{*}{Player A} & Strategy 1 ($C$) & $1,1$ & $-2,3$ \\
  & Strategy 2 ($D$) & $3,-2$ & $0,0$ \\
  \hline
\end{tabular}
}
\end{minipage}
\label{ts1}
\end{table}
\clearpage

\begin{table}[!t]
\caption{\textbf{Basic information on the conducted experimental sessions}. A total of seven sessions were conducted in the experiment, with considering both the probability of encountering zealots (denote as $q$) and the treatment conditions. Sessions
are characterized by the number of interactions, attendance, the mean age of participants and its standard deviation, and the percentage of female. The treatment conditions include the cross of without zealots (WZ), unaware of zealots (UoZ),
informed of zealots (IoZ), informed of zealot bots (IZB), and the probability of encountering zealots. Unless otherwise specified, the experiments are conducted using the payoff matrix I (PM I). The currency payment unit is CNY.
}
\label{tab_basicinfo}
\resizebox{\columnwidth}{!}{%
\begin{tabular}{cccccccccc}
\hline
Date & Session number& Participants &  Mean age (S.D.) & \% Female & \multicolumn{2}{c}{Experimental setup}  & Interactions & 
Mean monetary payoff [min, max] \\
\hline

\multirow{9}{*}{15, Oct, 2023}& \multirow{3}{*}{1} & \multirow{3}{*}{40} & \multirow{3}{*}{18.86 (0.82)}  & \multirow{3}{*}{50}  & \multirow{3}{*}{$q=1$ } & WZ  & 30 & \multirow{3}{*}{142.325 [92, 175]}  
\\ 
& & & &  & & UoZ & 30  &  \\ 
 & & & & & & IoZ  & 30  & \\ 
\cline{2-9}
 
  & \multirow{3}{*}{2} & \multirow{3}{*}{42} & \multirow{3}{*}{18.72 (0.69)}  & \multirow{3}{*}{50}  & \multirow{3}{*}{$q=1$ } & WZ  & 30 & \multirow{3}{*}{144.93 [101, 167]}  \\ 
& & & &  & & UoZ  & 30  &  \\ 
 & & & & & & IoZ & 30  & \\ 
 \cline{2-9}
 
  & \multirow{3}{*}{3} &\multirow{3}{*}{40} & \multirow{3}{*}{18.96 (0.8)}  & \multirow{3}{*}{50}  & \multirow{3}{*}{$q=1$ } & WZ & 30 & \multirow{3}{*}{143.4 [82, 180]}  \\ 
& & & &  & & UoZ  & 30  &  \\ 
 & & & & & & IoZ & 30  & \\ 
\hline

\multirow{9}{*}{21, Oct, 2023}& \multirow{3}{*}{4} & \multirow{3}{*}{40} & \multirow{3}{*}{18.85 (0.73)}  & \multirow{3}{*}{50}  & \multirow{3}{*}{$q=1$ (PM II) } & WZ & 30 & \multirow{3}{*}{115.05 [60, 137.5]}  \\ 
& & & &  & & UoZ  & 30  &  \\ 
 & & & & & & IoZ & 30  & \\ 
\cline{2-9}

  & \multirow{3}{*}{5} & \multirow{3}{*}{40} & \multirow{3}{*}{18.83 (0.86)}  & \multirow{3}{*}{50}  & \multirow{3}{*}{$q=1$ (PM II) } & WZ & 30& \multirow{3}{*}{116.325 [73, 136]}  \\ 
& & & &  & & UoZ  & 30  &  \\ 
 & & & & & & IoZ & 30  & \\ 
\cline{2-9}

  & \multirow{3}{*}{6} & \multirow{3}{*}{42} & \multirow{3}{*}{18.73 (0.59)}  & \multirow{3}{*}{47.6}  & \multirow{3}{*}{$q=1$ (PM II) } & WZ & 30 & \multirow{3}{*}{116.18 [67, 139]}  \\ 
& & & &  & & UoZ  & 30  &  \\ 
 & & & & & & IoZ & 30  & \\ 
\hline

\multirow{8}{*}{22, Oct, 2023} & \multirow{2}{*}{7} & \multirow{2}{*}{60} & \multirow{2}{*}{18.88 (0.73)}  & \multirow{2}{*}{50}  & \multirow{2}{*}{$q=0.5$} & UoZ & 15 & \multirow{2}{*}{69.415 [50, 84]}  \\ 
& & & &  &  & IoZ  & 15  &  \\ 
\cline{2-9}

  & \multirow{2}{*}{8} & \multirow{2}{*}{60} & \multirow{2}{*}{19.01 (0.66)}  & \multirow{2}{*}{50}  & \multirow{2}{*}{$q=0.5$} & UoZ & 15 & \multirow{2}{*}{69.15 [53, 88]}  \\ 
& &  & &  & & IoZ  & 15  &  \\ 
\cline{2-9}

    & \multirow{2}{*}{9} & \multirow{2}{*}{60} & \multirow{2}{*}{18.97 (0.7)}  & \multirow{2}{*}{50}  & \multirow{2}{*}{$q=0.25$} & UoZ & 15 & \multirow{2}{*}{54.535 [28, 70]}  \\ 
& & & &  & & IoZ  & 15  &  \\
\cline{2-9}

    & \multirow{2}{*}{10} & \multirow{2}{*}{60} & \multirow{2}{*}{18.79 (0.53)}  & \multirow{2}{*}{50}  & \multirow{2}{*}{$q=0.25$} & UoZ & 15 & \multirow{2}{*}{56.635 [36, 72]}  \\ 
& & & &  & & IoZ & 15  &  \\
\hline

\multirow{4}{*}{04, Nov, 2023} & \multirow{2}{*}{11} &  \multirow{2}{*}{60} & \multirow{2}{*}{18.82 (0.65)}  & \multirow{2}{*}{50}  & \multirow{2}{*}{$q=0.35$} & UoZ  & 15 & \multirow{2}{*}{61.415 [39, 80]}  \\ 
& & & &  &  & IoZ & 15  &  \\ 
\cline{2-9}

  & \multirow{2}{*}{12} & \multirow{2}{*}{60}  & \multirow{2}{*}{18.99 (1.17)}  & \multirow{2}{*}{50}  & \multirow{2}{*}{$q=0.35$} & UoZ & 15 & \multirow{2}{*}{60.9 [46, 76]}  \\ 
& & & &  & & IoZ & 15  &  \\ 
\hline

\multirow{4}{*}{05, Nov, 2023} & \multirow{2}{*}{13} &  \multirow{2}{*}{60} & \multirow{2}{*}{18.84 (0.97)}  & \multirow{2}{*}{50}  & \multirow{2}{*}{$q=0.75$} & UoZ & 15 & \multirow{2}{*}{79.035 [59, 92]}  \\ 
& & & &  &  & IoZ & 15  &  \\ 
\cline{2-9}

  & \multirow{2}{*}{14} & \multirow{2}{*}{60}  & \multirow{2}{*}{18.84 (0.56)}  & \multirow{2}{*}{50}  & \multirow{2}{*}{$q=0.75$} & UoZ & 15 & \multirow{2}{*}{80.865 [58, 98]}  \\ 
& & & &  & & IoZ & 15  &  \\ 
\hline

 11, Nov, 2023 & 15 & 40 & 18.83 (0.68) & 80 & $q=1$ & IZB  & 15 & 66.925 [55, 70] \\  
\hline
\multirow{2}{*}{18, Nov, 2023} & 16  & 40 & 18.87 (0.71) & 45 & \multirow{2}{*}{$q=1$} & \multirow{2}{*}{IZB}  & 15 & 67.475 [55, 70] \\  
   & 17 & 40 & 18.94 (0.57) & 25 & &   & 15  & 67.8 [58, 70] \\ 
\hline

\multirow{3}{*}{16, Mar, 2024} & 18 & 40 & 18.86 (0.69) & 65 & \multirow{3}{*}{$q=1$ (PM II)} & \multirow{3}{*}{IZB}  & 15 & 61.025 [55.5, 62.5] \\  
    & 19 & 40 & 19.21 (0.76) & 82.5 & &   & 15  & 60.9 [55.5, 62.5] \\ 
    & 20 & 40 & 19.07 (0.86) & 75 & &   & 15  & 60.625 [55.5, 62.5] \\
\hline
\end{tabular}%
}
\label{ts2}
\end{table}

\clearpage

\begin{table}[!t]
\centering
\caption{The table compares initial and final cooperation levels in treatments with participants unaware of zealots, informed of zealots, and informed of zealot bots versus treatment without zealots. It includes percentages and sample sizes for cooperation rates, alongside chi-square values ($\chi_{(1)}^2$) and $p$-values from two-sample proportion tests. All outcomes are derived under the context of payoff matrix I.}
\label{ts3}
\begin{tabularx}{\textwidth}{>{\hsize=1.5\hsize}X>{\hsize=0.5\hsize}Xcccc}
\toprule
\textbf{Treatment comparison} & \textbf{Stage} & \textbf{Cooperation rate (\%)} & \textbf{$\chi_{(1)}^2$} & \textbf{$p$-value} \\
\midrule
\multirow{2}{=}{Unaware of zealots vs. Without zealots} & Initial & 21.7 (53/244) vs. 27 (66/244) & 1.6 & 0.21 \\
 & Final & 31 (75/244) vs.\ 10.2 (25/244) & 30.2 & $<10^{-8}$ \\
\midrule
\multirow{2}{=}{Informed of zealots vs. Without zealots} & Initial & 29.7 (72/244) vs. 27 (66/244) & 0.253 & 0.62 \\
 & Final & 33.2 (81/244) vs.\ 10.2 (25/244) & 36.5 & $< 10^{-9}$ \\
\midrule
\multirow{2}{=}{Informed of zealot bots vs. Without zealots} & Initial & 18.3 (22/120) vs. 27 (66/244) & 2.88 & 0.09 \\
 & Final & 17.5 (21/120) vs.\ 10.2 (25/244) & 3.21 & 0.07 \\
\bottomrule
\end{tabularx}
\end{table}

\clearpage

\begin{table}[!t]
\renewcommand\arraystretch{1.1}
\setlength{\tabcolsep}{16pt}
\centering
\caption{Summary of Linear Mixed-Effects Models for treatments without zealots, unaware of zealots, and informed of zealots, conducted under payoff matrix I. The model investigates the impact of the sequence in which treatments are presented (referred to as 'Order') on the frequency of cooperation. 'Order' indicates the position of each treatment in the experimental schedule, treated as a categorical fixed effect that encompasses all sessions in a collective manner. The mixed-effects framework includes fixed effects to examine the influence of this treatment sequence and random effects to control for variability among individual participants. 'Order' is conceptualized as a sequence of distinct categories representing the progression of treatments, rather than as a numeric continuum. Random effects allow for adjustment based on individual participant differences. The findings reveal that the sequence of treatment exposure significantly affects cooperation frequencies in the without zealots and unaware of zealots conditions, suggesting that the specific timing of treatment presentation can modify participant behavior.}
\label{ts4}
\begin{tabular}{cccccc}
\toprule
Fixed effect & Estimate & Std. error & d.f.   & $t$ value & $p$ value      \\
\midrule
\multicolumn{6}{c}{\bf Without zealots} \\
\midrule
Intercept    & 0.21     & 0.03       & 143.93 & 7.21    & 2.89e-11 *** \\
Order 2      & -0.07    & 0.04       & 143.93 & -1.63   & 0.1          \\
Order 3      & -0.11    & 0.04       & 143.93 & -2.72   & 0.007 **     \\
Order 4      & -0.18    & 0.04       & 143.93 & -4.34   & 2.72e-05 *** \\
Order 5      & 0.04     & 0.04       & 143.93 & 0.86    & 0.39         \\
Order 6      & -0.07    & 0.02       & 3535   & -4.02   & 5.97e-05 *** \\
\midrule
\multicolumn{6}{c}{\bf Unaware of zealots} \\
\midrule
Intercept    & 0.39     & 0.05       & 125.5  & 7.16    & 6.13e-11 *** \\
Order 2      & -0.02    & 0.08       & 125.5  & -0.29   & 0.77         \\
Order 3      & -0.09    & 0.08       & 125.5  & -1.19   & 0.24         \\
Order 4      & -0.10    & 0.08       & 125.5  & -1.32   & 0.19         \\
Order 5      & -0.02    & 0.08       & 125.5  & -0.31   & 0.76         \\
Order 6      & -0.2     & 0.02       & 3535   & -11.4   & <2e-16 ***   \\
\midrule
\multicolumn{6}{c}{\bf Informed of zealots}  \\
\midrule
Intercept    & 0.32     & 0.06       & 124.2  & 5.46    & 2.51e-07 *** \\
Order 2      & -0.01    & 0.08       & 124.2  & -0.11   & 0.91         \\
Order 3      & 0.1      & 0.08       & 124.2  & 1.16    & 0.25         \\
Order 4      & 0.06     & 0.08       & 124.2  & 0.68    & 0.5          \\
Order 5      & -0.07    & 0.08       & 124.2  & -0.88   & 0.38         \\
Order 6      & -0.03    & 0.02       & 3535   & -1.86   & 0.06  \\
\bottomrule
\end{tabular}%
\end{table}

\clearpage

\begin{table}[!t]
\renewcommand\arraystretch{1.1}
\setlength{\tabcolsep}{8pt}
\centering
\caption{Participant distribution across SVO categories in a within-subject experiment under Payoff Matrix I. Following the classification guidelines proposed by ref. ~\cite{murphy2011measuring}, this table groups participants according to their SVO category—assessed across three treatments: without zealots (WZ), unaware of zealots (UoZ), and informed of zealots (IoZ). Each participant experienced all three conditions, thus the data for these treatments are aggregated. A separate group for the informed of zealot bots (IZB) treatment is also displayed. The table enumerates the participants in each SVO category and indicates the proportional representation within parentheses.}
\label{ts5}
\begin{tabular}{c|cccc}
\hline
\diagbox{Treatment type}{SVO category} & Competitive & Individualistic & Prosocial & Altruistic \\
\hline  
WZ \& UoZ \& IoZ & 6 (0.05) & 55 (0.45) & 60 (0.49) & 1 (0.008)  \\
IZB & 7 (0.06)   & 68 (0.57) & 45 (0.37) & 0 (0) \\
\hline
\end{tabular}
\end{table}

\clearpage
\begin{table}[!t]
\renewcommand\arraystretch{1.1}
\setlength{\tabcolsep}{17pt}
\centering
\caption{Results from the Analysis of Variance (ANOVA) assessing the impact of SVO categories (competitive, individualistic, prosocial) and treatment type (without zealots, unaware of zealots, informed of zealots, and informed of zealot bots) on cooperation levels. Although the main effects are significant, the interaction between SVO categories and treatment types does not show significance, suggesting that the relationship between SVO orientation and cooperation is not substantially different across the various treatments. For comprehensive post-hoc analysis details and further ANOVA results focusing on the impact of treatment types within each SVO category, refer to Supplementary Tables~\ref{ts7} and \ref{ts8}, respectively. All outcomes are derived under the context of payoff matrix I.}
\label{ts6}
%\resizebox{\columnwidth}{!}
\begin{tabular}{llllll}
\hline
Term               & d.f. & S.S   & MS   & $F$ statistic & $p$-value  \\ \hline
SVO                & 2    & 6.5   & 3.24 & 44.17       & \bf <2e-16   \\
Treatment          & 3    & 2.9   & 0.97 & 13.2        & \bf 2.78e-08 \\
SVO $\times$ Treatment     & 6    & 0.86  & 0.14 & 1.96        & 0.07     \\
Residuals          & 474  & 34.81 & 0.07 &             &          \\ \hline
Degrees of freedom &      &       &      &             &          \\
Sum of squares     &      &       &      &             &          \\
Mean squares       &      &       &      &             &          \\ \hline
\end{tabular}

\end{table}

\clearpage

\begin{table}[!t]
\renewcommand\arraystretch{1.1}
\setlength{\tabcolsep}{8pt}
\centering
\caption{Summary of Tukey Honest Significant Difference (HSD) post-hoc analysis for the impact of SVO categories (competitive, individualistic, and prosocial) and treatment types (without zealots, unaware of zealots, informed of zealots, and informed of zealot bots) on cooperation levels. The table presents differences in mean cooperation levels between each category and treatment, along with lower (LCB) and upper (UCB) confidence bounds, and associated $p$-values. Significant differences are noted between all treatments involving zealots and those without, as well as among different SVO categories, indicating a robust dependency of cooperation levels on both the treatment type and SVO orientation. All outcomes are derived under the context of payoff matrix I.}
\label{ts7}

\begin{tabular}{ccccccc}
\hline
\multicolumn{3}{c}{Comparison}          & Difference & LCB$^{a}$    & UCB$^{b}$    & $p$-value \\ \hline
Competitive     & vs. & Individualistic & 0.036      & -0.104 & 0.176  & 0.798   \\
Competitive     & vs. & Prosocial       & 0.264      & 0.124  & 0.403  & \bf 0.001   \\
Individualistic & vs. & Prosocial       & 0.228      & 0.166  & 0.289  & \bf 0.001   \\
Without zealots              & vs. & Informed of zealots              & 0.185      & 0.088  & 0.282  & \bf 0.001   \\
Without zealots              & vs. & Unaware of zealots              & 0.171      & 0.074  & 0.267  & \bf 0.001   \\
Without zealots              & vs. & Informed of zealot bots & 0.031      & -0.067 & 0.128  & 0.83    \\
Informed of zealots              & vs. & Unaware of zealots              & -0.014     & -0.111 & 0.083  & 0.9     \\
Informed of zealots              & vs. & Informed of zealot bots & -0.154     & -0.252 & -0.057 & \bf 0.001   \\
Unaware of zealots              & vs. & Informed of zealot bots & -0.14      & -0.237 & -0.043 & \bf 0.001   \\ \hline
\multicolumn{7}{l}{$^{a}$Lower confidence bound}                                       \\
\multicolumn{7}{l}{$^{b}$Upper confidence bound}                                       \\ \hline
\end{tabular}%
\end{table}

\clearpage

\begin{table}[!t]
\renewcommand\arraystretch{1.1}
\setlength{\tabcolsep}{8pt}
\centering
\caption{Analysis of Variance (ANOVA) examining how cooperation levels are influenced by treatment type (without zealots, unaware of zealots, informed of zealots, and informed of zealot bots) within SVO category (competitive, individualistic, and prosocial). The analysis reveals that treatment type significantly affects cooperation levels among prosocial individuals. In contrast, no such dependency is observed for competitive and individualistic individuals, suggesting that the impact of treatment type on cooperation is contingent upon the participants' SVO category. Detailed post-hoc comparisons for the effects of treatment types on cooperation levels within the prosocial category are presented in Supplementary Table~\ref{ts9}. All outcomes are derived under the context of payoff matrix I.}
\label{ts8}
\begin{tabular}{lllllll}
\hline
SVO category  & Term & d.f.$^{a}$ & SS$^{b}$ & MS$^{c}$ & $F$ statistics & $p$-value \\ \hline
\multirow{2}{*}{Competitive}     & Treatment type & 3    & 0.089  & 0.03   & 1.1          & 0.371             \\
                                 & Residuals      & 21   & 0.568  & 0.027  &              &                   \\
\multirow{2}{*}{Individualistic} & Treatment type & 3    & 0.298  & 0.994  & 2.583        & 0.054             \\
                                 & Residuals      & 229  & 8.811  & 0.0385 &              &                   \\
\multirow{2}{*}{Prosocial}       & Treatment type & 3    & 3.382  & 1.127  & 9.929        & \textbf{3.58e-06} \\
                                 & Residuals      & 224  & 25.428 & 0.1135 &              &                   \\ \hline
$^{a}$Degrees of freedom           &                &      &        &        &              &                   \\
$^{b}$Sum of squares             &                &      &        &        &              &                \\
$^{c}$Mean squares            &                &      &        &        &              &  \\ \hline
\end{tabular}%
\end{table}

\clearpage

\begin{table}[!t]
\renewcommand\arraystretch{1.1}
\setlength{\tabcolsep}{6pt}
\centering
\caption{Summary of the Tukey Honest Significant Difference (HSD) post-hoc test results assessing the effects of different treatment types (without zealots, unaware of zealots, informed of zealots, and informed of zealot bots) specifically on cooperation levels among prosocial individuals. This analysis corroborates the findings presented in Table~\ref{ts7}, where the influence of treatment types on cooperation levels was examined across all SVO categories collectively, underscoring consistent patterns of impact when focusing solely on prosocial individuals. All outcomes are derived under the context of payoff matrix I.}
\label{ts9}
\begin{tabular}{clccccc}
\hline
\multicolumn{3}{c}{Comparison}                 & Difference & LCB$^{a}$    & UCB$^{b}$    & $p$-value        \\ \hline
Without zealots                     & vs. & Informed of zealots              & 0.287      & 0.129  & 0.445  & \textbf{0.001} \\
Without zealots                      & vs. & Unaware of zealots              & 0.254      & 0.096  & 0.412  & \textbf{0.001} \\
Without zealots                      & vs. &Informed of zealot bots & 0.078      & -0.094 & 0.249  & 0.626          \\
Informed of zealots                     & vs. & Unaware of zealots              & -0.033     & -0.191 & 0.125  & 0.9            \\
Informed of zealots                     & vs. & Informed of zealot bots & -0.209     & -0.381 & -0.038 & \textbf{0.01}  \\
Unaware of zealots                     & vs. & Informed of zealot bots & -0.176     & -0.347 & -0.004 & \textbf{0.042} \\ \hline
$^a$Lower confidence bound &     &                 &            &        &        &                \\
$^b$Upper confidence bound &     &                 &            &        &        &                \\ \hline
\end{tabular}%
\end{table}

\clearpage
\begin{table}[!t]
\renewcommand\arraystretch{1.1}
\setlength{\tabcolsep}{8pt}
\centering
\caption{Summary of participant distribution across SVO categories using the slide measure in a within-subject design, under payoff matrix II. Participants engaged in three treatments—without zealots, unaware of zealots, and informed of zealots—are grouped together as they encountered each condition. The table also presents data for a separate group under the informed of zealot bots treatment. The numbers reflect the count of participants in each category, with percentages provided in parentheses.}
\label{ts10}
\begin{tabular}{c|cccc}
\hline
\multicolumn{1}{c|}{\diagbox{Treatment type}{SVO category}} & Competitive & Individualistic & Prosocial & Altruistic \\ 
\hline  
WZ \& UoZ \& IoZ & 1 (0.008) &  44 (0.36) &  77 (0.63) & 0 (0) \\
IZB & 9 (0.075)   & 74 (0.62) & 36 (0.3) & 1 (0.008) \\
\hline
\end{tabular}
\end{table}

\clearpage

\begin{table}[!t]
\centering
\caption{The table compares initial and final cooperation levels in treatments with participants unaware of zealots, informed of zealots, and informed of zealot bots versus treatment without zealots. It includes percentages and sample sizes for cooperation rates, alongside chi-square values ($\chi_{(1)}^2$) and $p$-values from two-sample proportion tests.  All outcomes are derived under the context of payoff matrix II.}
\label{ts11}
\begin{tabularx}{\textwidth}{>{\hsize=1.5\hsize}X>{\hsize=0.5\hsize}Xcccc}
\toprule
\textbf{Treatment comparison} & \textbf{Stage} & \textbf{Cooperation rate (\%)} & \textbf{$\chi_{(1)}^2$} & \textbf{$p$-value} \\
\midrule
\multirow{2}{=}{Unaware of zealots vs. Without zealots} & Initial & 15.6 (38/244) vs. 17.5 (43/244) & 0.24 & 0.63 \\
 & Final & 25.4 (62/244) vs.\ 4 (10/244) & 42.4 & $<10^{-11}$ \\
\midrule
\multirow{2}{=}{Informed of zealots vs. Without zealots} & Initial & 25.8 (63/244) vs. 17.5 (43/244) & 4.35 & 0.04 \\
 & Final & 27.5 (67/244) vs.\ 4 (10/244) & 48.4 & $< 10^{-12}$ \\
\midrule
\multirow{2}{=}{Informed of zealot bots vs. Without zealots} & Initial & 19.2 (23/120) vs. 17.5 (43/244) & 0.05 & 0.83 \\
 & Final & 13.3 (16/120) vs.\ 4 (10/244) & 8.997 & 0.003 \\
\bottomrule
\end{tabularx}
\end{table}

\clearpage

\begin{table}[!t]
\renewcommand\arraystretch{1.1}
\setlength{\tabcolsep}{15pt}
\centering
\caption{Summary of linear mixed-effects models for treatments without zealots, unaware of zealots, and informed of zealots under payoff matrix II. The definition of order and the employed linear mixed-effects model are consistent with those in Table~\ref{ts4}.}
\label{ts12}
\begin{tabular}{cccccc}
\toprule
Fixed effect & Estimate & Std. error & d.f.   & $t$ value & $p$ value      \\
\midrule
\multicolumn{6}{c}{\bf Without zealots} \\
\midrule
Intercept    & 0.13     & 0.02       & 159.44 & 6.52    & 8.68e-10 *** \\
Order 2      & -0.03    & 0.03       & 159.44 & -1.08   & 0.28          \\
Order 3      & -0.05    & 0.03       & 159.44 & -1.72   & 0.08     \\
Order 4      & -0.11    & 0.03       & 159.44 & -3.96   & 0.0001 *** \\
Order 5      & -0.04     & 0.03       & 159.44 & -1.48    & 0.14         \\
Order 6      & -0.04    & 0.01       & 3535   & -2.94   & 0.003 ** \\
\midrule
\multicolumn{6}{c}{\bf Unaware of zealots} \\
\midrule
Intercept    & 0.26     & 0.05       & 126.0  & 5.09    & 1.29e-06 *** \\
Order 2      & 0.02    & 0.07       & 126.0  & 0.23   & 0.82         \\
Order 3      & -0.004    & 0.07       & 126.0  & -0.06   & 0.95         \\
Order 4      & -0.01    & 0.07       & 126.0  & 0.14   & 0.89         \\
Order 5      & -0.03    & 0.07       & 126.0  & -0.35   & 0.73         \\
Order 6      & -0.1     & 0.02       & 3535   & -5.88   & <4.52e-09 ***   \\
\midrule
\multicolumn{6}{c}{\bf Informed of zealots}  \\
\midrule
Intercept    & 0.25     & 0.05       & 124.29  & 4.66    & 7.98e-06 *** \\
Order 2      & 0.07    & 0.08       & 124.29  & 0.85   & 0.4         \\
Order 3      & 0.07      & 0.08       & 124.29  & 0.85    & 0.4         \\
Order 4      & 0.02     & 0.08       & 124.29  & 0.29    & 0.77          \\
Order 5      & -0.03    & 0.08       & 124.29  & -0.37   & 0.71         \\
Order 6      & -0.01    & 0.02       & 3535   & -0.89   & 0.37  \\
\bottomrule
\end{tabular}%
\end{table}

\clearpage

\begin{table}[]
\renewcommand\arraystretch{1.1}
\setlength{\tabcolsep}{3pt}
\caption{Distribution and Chi-Square Analysis of SVO Categories Across Treatment Conditions. The table delineates the distribution of participants across SVO categories—Competitive, Individualistic, and Prosocial/Altruistic—in various treatments: WZ (Without Zealots), UoZ (Unaware of Zealots), and IoZ (Informed of Zealots), at different probabilities of encountering zealots ($q$). The numbers represent observed frequencies with their proportions in parentheses. Chi-square test statistics are provided to assess the distribution differences among the treatments. The results indicate no significant distribution variation across SVO categories, uggesting that the probability of encountering zealots does not significantly alter the distribution of SVO types among participants in these conditions.}
\begin{tabular}{ccccc}
\hline
\multicolumn{1}{c|}{\diagbox{Treatment type}{SVO category}} &
  Competitive &
  Individualistic &
  \multicolumn{1}{c|}{Prosocial/Altruistic} &
  \multicolumn{1}{l}{$\chi^2$ test result} \\ \hline
\multicolumn{1}{c|}{WZ \& UoZ \& IoZ} &
  6 (0.05) &  55 (0.45) &   \multicolumn{1}{c|}{61 (0.50)} &
  \multirow{6}{*}{\begin{tabular}[c]{@{}c@{}}$\chi^2$=13.44, \\ ${d.f.}^{a}=10$, \\ p-value=0.2\end{tabular}} \\
\multicolumn{1}{c|}{UoZ ($q=0.75$) \& IoZ ($q=0.75$)} & 3 (0.03) & 64 (0.53) & \multicolumn{1}{c|}{53 (0.44)} &                      \\
\multicolumn{1}{c|}{UoZ ($q=0.5$) \& IoZ ($q=0.5$)}   & 5 (0.04) & 57 (0.47) & \multicolumn{1}{c|}{58 (0.49)} &                      \\
\multicolumn{1}{c|}{UoZ ($q=0.35$) \& IoZ ($q=0.35$)} & 5 (0.04)  & 50 (0.42) & \multicolumn{1}{c|}{65 (0.54)} &                      \\
\multicolumn{1}{c|}{UoZ ($q=0.25$) \& IoZ ($q=0.25$)} & 5 (0.04) & 52 (0.43) & \multicolumn{1}{c|}{63 (0.53) } &                        \\ \hline
\multicolumn{4}{l}{$^{a}$degree of freedom}                                                & \multicolumn{1}{l}{} \\ \hline
\end{tabular}
\label{ts13}
\end{table}

\clearpage

\begin{table}[!t]
\small
\caption{The threshold effect of zealots in promoting cooperation in scenarios of unaware of zealots and informed of zealots, as determined by a generalized linear regression model. The model assesses the impact of the probability of encountering zealots on participants' cooperation levels, with the without zealots scenario and competitive individuals serving as the baseline group. Social value orientation is controlled for in the analysis. The generalized linear regression model indicates that a minority of zealots can enhance cooperation, particularly when players are informed about their opponents' zealotry. Regarding the impact of social value orientations on cooperation in the whole experiment, the generalized linear regression model indicates that individualistic individuals exhibit a higher probability of cooperation only in the informed of zealots scenario compared to competitive individuals, whereas prosocial individuals show significantly greater cooperation levels in both scenarios. }
\label{ts14}
\resizebox{\columnwidth}{!}{
\begin{tabular}{ccccccccccc}
\hline
\multirow{2}{*}{}      & \multicolumn{10}{c}{\multirow{2}{*}{\begin{tabular}[c]{@{}c@{}}Dependent variable: participants whether to cooperate \\ in each round of games\end{tabular}}} \\
    \multicolumn{10}{c}{}  \\ \hline
\multirow{2}{*}{Model} & \multicolumn{4}{c}{Unaware of zealots }                  &        &        & \multicolumn{4}{c}{Informed of zealots }                 \\ \cline{2-11} 
 & $Coef.^{a}$       & $S.E.^{b}$       & $z$ value       & pr(\textgreater{}|$z$|)       &        &        & $Coef.^{a}$     & $S.E.^{b}$      & $z$ value      & pr(\textgreater{}|$z$|)      \\ \hline
Constant               & -2.56      & 0.14       & -18.52         & \textless{} 2e-16 ***           &        &        & -2.83      & 0.14      & -19.68        & \textless{} 2e-16 ***           \\
$q=0.25$               & -0.32       & 0.09       & -3.52        & 0.0004 ***                    &        &        & 0.17       & 0.08      & 2.07         & 0.04*                       \\
$q=0.35$               & -0.08       & 0.08       & -0.96          & 0.34                        &        &        & 0.31       & 0.08      & 3.97         & 7.08e-05 ***                      \\
$q=0.5$               & 0.26        & 0.08       & 3.24         & 0.001 **                        &        &        & 0.54       & 0.08      & 7.09         & 1.37e-12 ***                   \\
$q=0.75$              & 0.51        & 0.08       & 6.67          & 4.49e-11 ***                    &        &        & 0.72       & 0.08      & 9.62         & \textless{} 2e-16 ***                      \\
$q=1$                 & 1.06        & 0.06       & 17.5          & \textless{} 2e-16 ***           &        &        & 1.14       & 0.06      & 18.75        & \textless{}2e-16 ***          \\
Individualistic        & 0.096        & 0.14       & 0.71          & 0.48                        &        &        & 0.36       & 0.14      & 2.51         & 0.01 ***                   \\
Prosocial             & 1.21        & 0.13       & 9.06          & \textless{} 2e-16 ***                      &        &        & 1.51       & 0.14      & 10.8         & \textless{} 2e-16 ***                   \\
Null deviance          & \multicolumn{4}{c}{14233}                                              &        &        & \multicolumn{4}{c}{15260}                                          \\
Residual deviance      & \multicolumn{4}{c}{13102}                                              &        &        & \multicolumn{4}{c}{14071}                                          \\
AIC$^{c}$                    & \multicolumn{4}{c}{13118}                                              &        &        & \multicolumn{4}{c}{14087}                                          \\
Observation            & \multicolumn{4}{c}{14520}                                              &        &        & \multicolumn{4}{c}{14520}                                          \\ \hline
\multicolumn{11}{l}{$^{a}$Coefficient}\\
\multicolumn{11}{l}{$^{b}$Standard error} \\
\multicolumn{11}{l}{$^{c}$Akaike information criterion}                                                                       \\ \hline
\end{tabular}
}
\end{table}

\clearpage

\begin{sidewaystable}[]
\caption{\textbf{Contingency tables for gender as a confounding factor}. With regard to the probabilities of encountering zealots (denoted as $q$), the contingency tables for each treatment are presented by the following information: the observed cell totals, (the expected cell totals), and [the $\chi^2$ statistic for each cell]. The ``informed of zealot bots'' are denoted as IZB. Unless otherwise specified, the experiments are conducted using the payoff matrix I (PM I). At a 95\% confidence level, the critical value of the chi-square distribution with 1 degree of freedom is 3.84, and contingency tables with statistically significant differences are marked with an asterisk (*).}
\label{tab_gender}
\resizebox{\textwidth}{!}{%

\begin{tabular}{cccccccccccccc}
\hline
& \multicolumn{9}{c}{\textbf{Treatment condition}} \\ 
\multirow{2}{*}{Encountering probability} & \multirow{2}{*}{Gender} & \multicolumn{2}{c}{Without zealots} & & \multicolumn{2}{c}{Unaware of zealots} & & \multicolumn{2}{c}{Informed of zealots} & & \multicolumn{2}{c}{Informed of zealot bots} \\ 
\cline{3-4} \cline{6-7} \cline{9-10} \cline{12-13}
& & Cooperation & Defection & & Cooperation & Defection & & Cooperation & Defection & & Cooperation & Defection  \\
\hline
\multirow{2}{*}{$q=1$}& Female & 214 (261)[8.464] * & 1616 (1569)[1.408] * & & 507 (573.5)[7.711] * & 1323 (1256.5)[3.52] * & & 456 (599.5)[34.35] *  & 1374 (1230.5)[16.734] *  & & - & -  \\

& Male & 308 (261)[8.464] * & 1522 (1569)[1.408] * & & 640 (573.5)[7.711] *  & 1190 (1256.5)[3.52] *  & & 734 (599.5)[34.35] *  & 1087 (1230.5)[16.734] *  & & - & - \\

\multirow{2}{*}{$q=1$ (PM II)}& Female & 145 (151.97)[0.319] & 1655 (1648.03)[0.029] & & 452 (434.75)[0.684] & 1348 (1365.25)[0.218] & & 512 (489.84)[1.003] & 1288 (1310.16)[0.375] & & - & - \\

& Male & 164 (157.03)[0.309] & 1696 (1702.97)[0.029] & & 432 (449.25)[0.662] & 1428 (1410.75)[0.211] & & 484 (506.16)[0.971] & 1376 (1353.84)[0.363] & & - & - \\

\multirow{2}{*}{$q=1$ (IZB)}& Female & - & - & & - & - & & - & - &  & 134 (156)[3.103] * & 766 (744)[0.651] *  \\

& Male & - & - & & - & - & & - & - & & 178 (156)[3.103] * & 722 (744)[0.651] *   \\

\multirow{2}{*}{$q=1$ (IZB, PM II)}& Female & - & - &  & - & - &  & - & - & & 130 (146.85)[4.333] * & 1205 (1188.15)[285.64] *  \\

& Male & - & - &  & - & - & & - & - & & 68 (51.15)[49.64] * & 397 (413.85)[161.84] *   \\

\multirow{2}{*}{$q=0.75$ }& Female & - & - & & 125 (183.5)[18.65] * & 775 (716.5)[4.78] * & & 154 (213.5)[16.582] * & 746 (686.5)[5.157] * & & - & - \\

& Male & - & - & & 242 (183.5)[18.65] * & 658 (716.5)[4.776] * & & 273 (213.5)[16.582] * & 627 (686.5)[5.517] * & & - & - \\

\multirow{2}{*}{$q=0.5$ }& Female & - & - & & 181 (156)[4.006] * & 719 (744)[0.84] * & & 207 (193.5)[0.942] & 693 (706.5)[0.258] & \\
& Male & - & - & & 131 (156)[4.006] * & 769 (744)[0.84] * & & 180 (193.5)[0.942] & 720 (706.5)[0.258] & & - & - \\

\multirow{2}{*}{$q=0.35$ }& Female & - & - & & 114 (124.5)[0.886] & 786 (775.5)[0.142] & & 143 (171)[4.585] * & 757 (729)[1.075] * & \\
& Male & - & - & & 135 (124.5)[0.886] & 765 (775.5)[0.142] & & 199 (171)[4.585] * & 701 (729)[1.075] * & & - & - \\

\multirow{2}{*}{$q=0.25$ }& Female & - & - & & 81 (100.5)[3.784] * & 819 (799.5)[0.476] * & & 123 (150.5)[5.025] * & 777 (749.5)[1.009] *  \\
& Male & - & - & & 120 (100.5)[3.784] * & 780 (799.5)[0.476] * & & 178 (150.5)[5.025] * & 722 (749.5)[1.009] * & & - & - \\
\hline
\end{tabular}
}
\end{sidewaystable}

\clearpage

\begin{sidewaystable}[]
\caption{\textbf{Contingency tables for academic background as a confounding factor.} With regard to the probabilities of encountering zealots (denoted as $q$), the contingency tables for each treatment are presented by the following information: the observed cell totals, (the expected cell totals), and [the $\chi^2$ statistic for each cell]. The ``informed of zealot bots'' denote as IZB. The participants are divided into two groups based on their major field of study, namely mathematics and natural sciences (denoted as M\&NS) and humanities and social sciences (denoted as H\&SS). Unless otherwise specified, the experiments are conducted using the payoff matrix I(PM I). At a 95\% confidence level, the critical value of the chi-square distribution with 1 degree of freedom is 3.84, and contingency tables with statistically significant differences are marked with an asterisk (*).
}
\label{tab_major}
\resizebox{\textwidth}{!}{
\begin{tabular}{cccccccccccccc}
\hline
& \multicolumn{9}{c}{\textbf{Treatment condition}} \\ 
\multirow{2}{*}{Encountering probability} & \multirow{2}{*}{Gender} & \multicolumn{2}{c}{Without zealots} & & \multicolumn{2}{c}{Unaware of zealots} & & \multicolumn{2}{c}{Informed of zealots} & & \multicolumn{2}{c}{Informed of zealot bots} \\ 
\cline{3-4} \cline{6-7} \cline{9-10} \cline{12-13}
& & Cooperation & Defection & & Cooperation & Defection & & Cooperation & Defection & & Cooperation & Defection  \\ 
\\
\hline

\multirow{2}{*}{$q=1$}& M\&NS &  343 (372.25)[2.298] * &  2267(2237.75)[0.382] * & & 820 (817.93)[0.005] & 1790 (1792.06)[0.002] & & 931 (855.02)[6.751] * & 1679 (1754.98)[3.289] *  \\
& H\&SS & 179 (149.75)[5.712] * & 871 (900.25)[0.95] * & & 327 (329.06)[0.013] & 723 (720.94)[0.006] & & 268 (343.98)[16.781] * & 782 (706.02)[8.176] * & & - & -  \\

\multirow{2}{*}{$q=1$ (PM II)}& M\&NS &  127 (119.04)[0.532] &  1283 (1290.96)[0.049] & & 333 (340.56)[0.168] & 1077 (543.44)[0.053] & & 449 (383.7)[11.111] * & 961 (1026.3)[4.154] *  \\
& H\&SS & 182(189.96)[0.333] & 2068 (2060.04)[0.03] & & 551 (543.44)[0.105] & 1699 (1706.56)[0.033] & & 547 (612.3)[6.963] * & 1703 (1637.71)[2.603] * & & - & -  \\

\multirow{2}{*}{$q=1$ (IZB) }& M\&NS & - & - & & - & - & & - & - &  & 79 (82.32)[0.134] & 416 (412.68)[0.027]  \\
& H\&SS & - & - & & - & - &  & - & - &  &  153 (149.58)[0.074] & 747 (750.32)[0.015]   \\

\multirow{2}{*}{$q=1$ (IZB, PM II) }& M\&NS & - & - & & - & - &  & - & - & & 65 (61.05)[0.256] & 490 (493.95)[0.032]  \\
& H\&SS & - & - & & - & - & & - & - & & 133 (136.95)[0.114] & 1112 (1108.05)[0.014]   \\

\multirow{2}{*}{$q=0.75$ }& M\&NS & - & - & & 0 (3.06)[3.058] & 15 (11.94)[0.783] & & 6 (3.56)[1.675] & 9 (11.44)[0.521]  \\
& H\&SS & - & - & & 367 (363.94)[0.026] & 1418 (1421.06)[0.007] & & 421 (423.44)[0.014] & 1364 (1361.56)[0.004] & & - & -   \\

\multirow{2}{*}{$q=0.5$ }& M\&NS & - & - & & 94 (72.28)[6.525] * & 266 (287.72)[1.639] * & & 105 (83.86)[5.33] * & 255 (276.14)[1.619] *  \\
& H\&SS & - & - & & 162 (183.72)[2.567] * & 753(731.28)[0.645] * & & 192 (213.14)[2.097] * & 723 (701.86)[0.637] * & & - & -  \\

\multirow{2}{*}{$q=0.35$ }& M\&NS & - & - & & 2 (2.35)[0.052] & 15 (14.65)[0.008] & & 6 (6.82)[0.099] & 30 (29.18)[0.023]  \\
& H\&SS & - & - & & 249 (248.65)[0.0004] & 1551 (1551.35)[0.000] & & 342 (341.18)[0.001] & 1458 (1458.82)[0.000]  & & - & -  \\

\multirow{2}{*}{$q=0.25$ }& M\&NS & - & - & & 44 (51.39)[1.062] & 511 (503.61)[0.108] & & 80 (85.1)[0.306] & 475 (469.9)[0.055]  \\
  & H\&SS & - & - & & 81 (73.61)[0.742] & 714 (721.39)[0.076] & & 127 (121.9)[0.213] & 668 (673.1)[0.039] & & - & -  \\ 
\hline
\end{tabular}
}
\end{sidewaystable}
\clearpage

% \bibliographystyle{elsarticle-num}
% \bibliography{Ref-SI}

%\end{document}

\end{document}